\documentclass[sigconf,nonacm]{acmart}
\usepackage{braket}
\usepackage{amsmath}
\usepackage{graphicx}
\usepackage{subcaption}
\usepackage{tikz}
\usetikzlibrary{quantikz}
\usepackage{adjustbox}

\AtBeginDocument{%
  \providecommand\BibTeX{{%
    \normalfont B\kern-0.5em{\scshape i\kern-0.25em b}\kern-0.8em\TeX}}}

\begin{document}

\title{Hardware Efficient Quantum Search Algorithm}

\author{Ji Liu}
\email{jliu45@ncsu.edu}
\orcid{0000-0002-5509-5065}
\affiliation{%
  \institution{North Carolina State University}
  \city{Raleigh}
  \state{North Carolina}
  \country{USA}
  \postcode{27606}
}

\author{Huiyang Zhou}
\email{hzhou@ncsu.edu}
\orcid{0000-0003-2133-0722}
\affiliation{%
  \institution{North Carolina State University}
  \city{Raleigh}
  \state{North Carolina}
  \country{USA}
  \postcode{27606}
}

\begin{abstract}
Quantum computing has noteworthy speedup over classical computing by taking advantage of quantum parallelism, i.e., the superposition of states. In particular, quantum search is widely used in various computationally hard problems. Grover's search algorithm finds the target element in an unsorted database with quadratic speedup than classical search and has been proved to be optimal in terms of the number of queries to the database. The challenge, however, is that Grover's search algorithm leads to high numbers of quantum gates, which make it infeasible for the Noise-Intermediate-Scale-Quantum (NISQ) computers. 

In this paper, we propose a novel hardware efficient quantum search algorithm to overcome this challenge. Our key idea is to replace the global diffusion operation with low-cost local diffusions. Our analysis shows that our algorithm has similar oracle complexity to the original Grover's search algorithm while significantly reduces the circuit depth and gate count. The circuit cost reduction leads to a remarkable improvement in the system success rates, paving the way for quantum search on NISQ machines.
\end{abstract}




\maketitle

\section{Introduction}
Quantum computing has become an emerging field due to its potential speedup over classical computing. Recently, Google and IBM have announced their superconducting quantum computers with 72 and 65 qubits~\cite{Google72qubit,IBM65qubit}. Honeywell also announced its 10 qubits trapped-ion quantum computer~\cite{Honeywell10qubit}. These quantum computers with few tens to few hundreds of qubits are termed as Noisy Intermediate-Scale Quantum (NISQ) computers~\cite{preskill2018nisq}.

When executing on the NISQ computers, quantum algorithms need to be optimized based on the property of the target hardware. Many famous quantum algorithms were invented last century and are typically assumed to be executed on a fully connected quantum computer. However, state-of-the-art quantum computers usually have limited connectivity. The quantum circuits of the algorithm need to be compiled based on the coupling map of the target hardware. The compiler will insert SWAP gates and perform optimization to ensure the resulting circuit fits the target hardware. This compilation step might lead to a non-negligible increment to the circuit depth and the total number of quantum gates. Hence, there is an emerging need to develop hardware efficient quantum algorithms. 

Grover's search algorithm~\cite{grover1996fast} was invented in 1996. It finds the target item in a database. Grover's algorithm has been proven to be optimal in oracle complexity, in other words, the number of queries to the database. For a database with $N$ items, a classical search algorithm needs $O(N)$ queries to find the target item. On the other hand, Grover's algorithm only needs $O(\sqrt{N})$ queries.
Grover's algorithm is the only threat to postquantum cryptography~\cite{zhang2020partialgroverdepth}. In the postquantum cryptography standard, NIST suggests an approach where quantum attacks are restricted a fixed circuit depth~\cite{NISTpostquantum}. Thus, it is important to perform depth optimization for the quantum algorithms. Grover's algorithm consists of two units, the oracle, and the diffusion operator. The oracle flips the phase of the target item, and the diffusion operator amplifies the probability amplitude of the target item. By iteratively applying the oracle and the diffusion operator, the probability of finding the target item will approach $100\%$.

In this paper, we proposed a hardware efficient quantum search algorithm. Our algorithm can be viewed as a sequence of local Grover searches. For example, an $n$ qubit search problem can be divided into two local searches with $m$ and $n-m$ qubits, respectively. The first local search will amplify the items for which the leading $m$ qubits are the same as the target item. Then the second local search amplifies the items that have the following $n-m$ qubits in the correct state. By combining these two local searches, our algorithm will find the target item with high probability. Our algorithm has the same oracle complexity $O(\sqrt{N})$ as the naive Grover's algorithm. However, since we split the $n$ qubit diffusion operator into several local operators, the circuit depth and the number of quantum gates have been significantly reduced. Moreover, local operators are more favorable on the NISQ computers since they limit the noise from propagating from one set of qubits to another. Our experimental results show that our hardware efficient design leads to an $85\%$ improvement in the success rate. For the quantum computers with limited connectivity, we can partition the operator based on the coupling map of the computer. The local diffusion operators can be placed on the region with high qubit connectivity. 

\begin{figure*}[ht]
\centering 
    \begin{subfigure}[t]{0.32\textwidth}
        \includegraphics[width=\columnwidth]{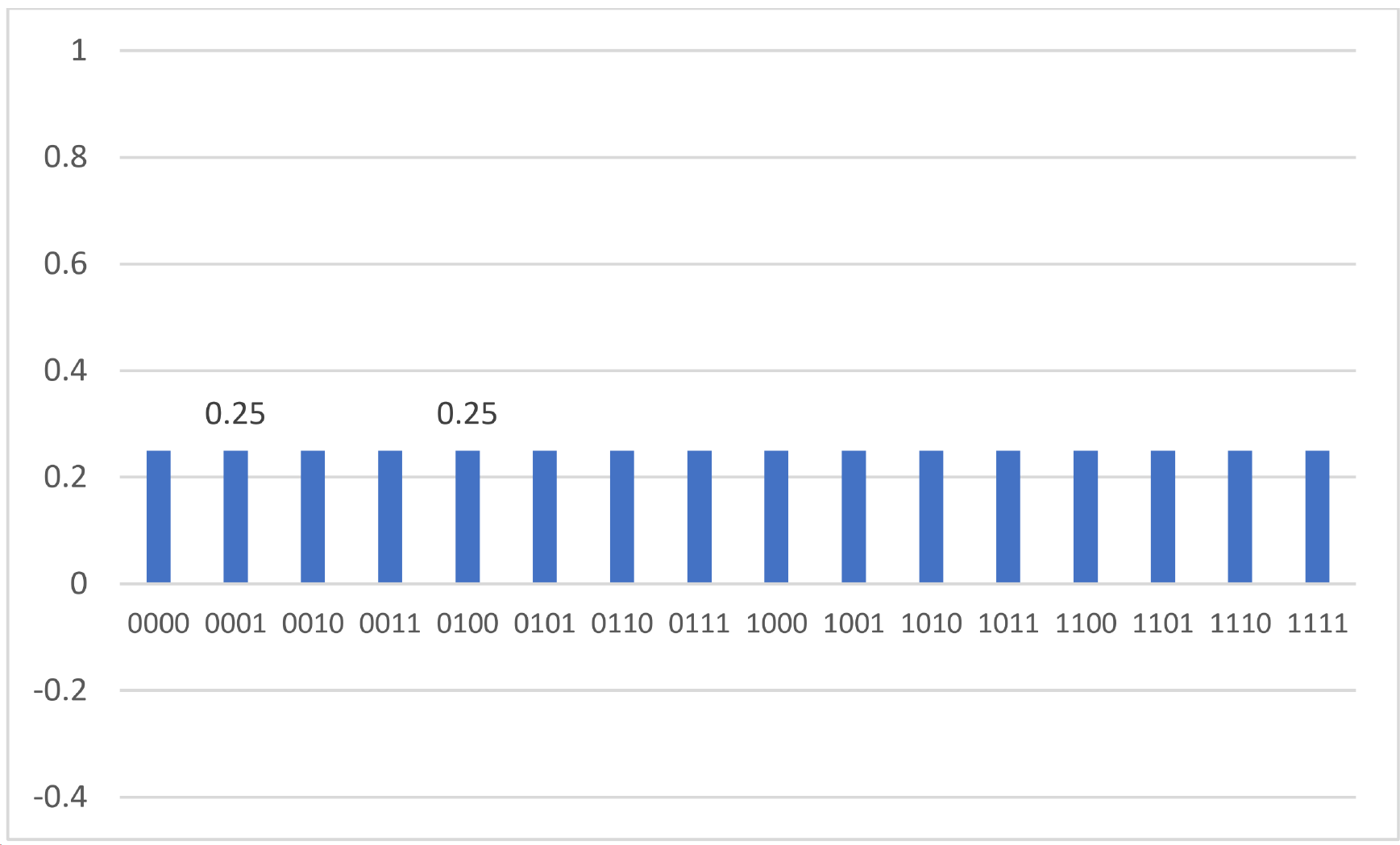}
        \caption{The initial quantum state is an equally distributed superposition state of all items.}
    \end{subfigure}\hfill
    \begin{subfigure}[t]{0.32\textwidth}
        \includegraphics[width=\columnwidth]{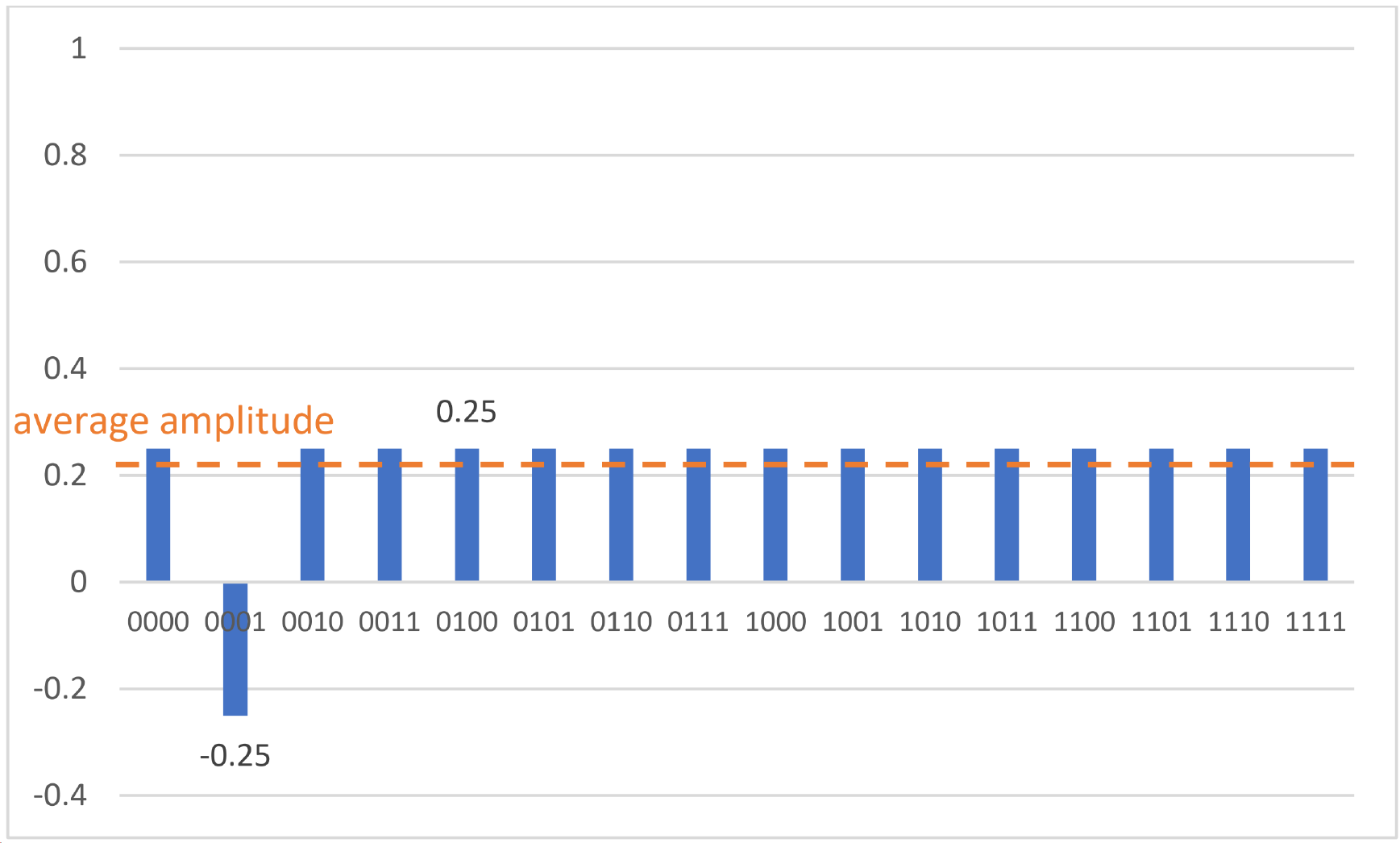}
        \caption{The oracle flips the probability amplitude of the target item, leads to a decrease in the average amplitude.}
    \end{subfigure}\hfill
        \begin{subfigure}[t]{0.32\textwidth}
        \includegraphics[width=\columnwidth]{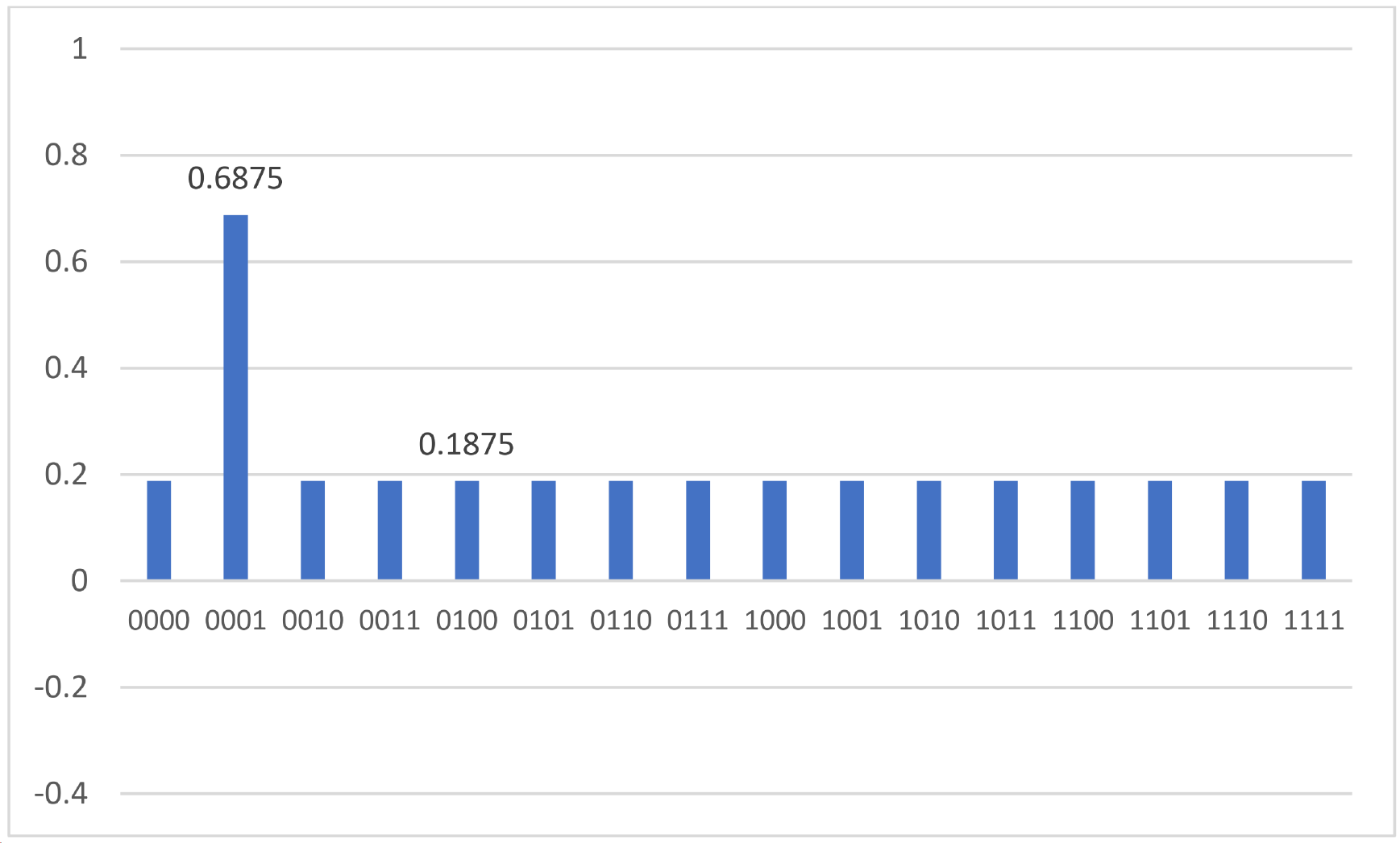}
        \caption{The diffusion operator inverts the amplitudes about average, which amplifies the amplitude of the target item.}
    \end{subfigure}\hfill
\caption{The first iteration of the 4-qubit Grover's algorithm}
\label{fig:naive_grover_illu}
\end{figure*}

The major contributions of our paper are as follows:
\begin{itemize}
\item We propose a hardware efficient quantum search algorithm.
\item We analyze the oracle and depth complexity of our proposed algorithm.
\item We discuss the design trade-off and how to choose the correct parameters.
\item We show that our proposed algorithm achieves less circuit depth and fewer quantum gates than the state-of-the-art quantum search algorithms.
\end{itemize}

The remainder of the paper is organized as follows: Section~\ref{sec:background} introduces the background of quantum computing, Grover's search algorithm, and the partial search algorithm. Section~\ref{sec:hardware_efficient_search} describes our hardware efficient quantum search algorithm and derives the matrix representation of our algorithm. Section~\ref{sec:hardware_efficient_search} also discusses the design trade-off and how to choose correct parameters. Section~\ref{sec:methodology} presents our experimental methodology. Section~\ref{sec:results} compares our algorithm with the existing quantum search algorithms. Section~\ref{sec:conclusion} summarizes the paper.

\section{Background And Related Work}
\label{sec:background}
\subsection{Quantum Computing}
Similar to bit in classical computing, qubit(quantum bit) is the basic unit in classical computing. A quantum bit can stay in the superposition of two classical states $\ket{0}$ and $\ket{1}$. The superposition state can be represented as $\ket{\psi} = a\ket{0} + b\ket{1}$, where $a$ and $b$ are complex numbers and $\left | a \right |^2 + \left | b \right |^2= 1$. Here $a$ and $b$ are the probability amplitudes of state $\ket{0}$ and $\ket{1}$. When measuring the superposition state, the probabilities of getting $\ket{0}$ and $\ket{1}$ states are $\left | a \right |^2$ and $\left | b \right |^2$, respectively. The superposition indicates implicit parallelism of quantum states which is critical to the speed up of quantum algorithms. The state of a quantum system is represented by a vector $\ket{\psi}$~\cite{nielsen2002quantumcomputation}. When multiple qubits are not entangled, the state of the multiple qubits can be expressed as the tensor product of each qubit state: $\ket{\psi_{12}} = \ket{\psi_1}\otimes\ket{\psi_2} = \ket{\psi_1}\ket{\psi_2}$. Entanglement means the measurement results of individual qubits are correlated. For example, the state $\frac{1}{\sqrt{2}}(\ket{00} + \ket{11})$ is an entangled state. After measurement, if the first qubit is in $\ket{0}$ state, the second qubit is in $\ket{0}$ state as well. 

A quantum program consists of a sequence of quantum gates. Quantum gates can be represented with unitary matrices. The quantum gate operates on a quantum state can be expressed as matrix-vector multiplications. For example, when an n-qubit quantum gate operates on a n-qubit state, the output state can be expressed as a $2^n\times2^n$ matrix $U$ multiplies a $2^n$ vector ${\ket{\psi_{input}}}$: $\ket{\psi_{output}} = U\ket{\psi_{input}}$. A sequence of quantum gates can be viewed as a single quantum gate with the corresponding matrix:$U = U_1U_2...U_n$.

\subsection{Grover's Algorithm}

Grover's algorithm finds the target item in a database. Grover's algorithm prepares the qubits in an equally distributed state and applies the Grover operator $G$ for $k$ times. The Grover operator $G$ consists of an oracle $O$ and a diffusion operator $D$. The circuit of 4-qubit Grover's algorithm is shown in Figure~\ref{fig:naivegrovercircuit}. The oracle is designed such that it will flip the probability amplitude of the target items. The action of oracle $O$ may be written as:
\begin{equation} 
\ket{x}\overset{O}{\rightarrow}(-1)^{f(x)}\ket{x}
\end{equation}
Here, $f(x)$ is the function of the database query which returns $1$ for the target items and $0$ for the non-target items. The second operator is the Grover's diffusion operator $D$, it inverts the amplitude of quantum states about average amplitude. An n-qubit diffusion operator can be expressed as:
\begin{equation} 
D = 2\ket{\psi}\bra{\psi} - I^{\otimes n}
\end{equation}
where $\ket{\psi}$ is the normalized sum of equally distributed superposition states.
\begin{figure}[ht]
\centering
\begin{adjustbox}{width=\linewidth}
\begin{quantikz}
\lstick{$q_0$} & \gate[wires=4]{\begin{array}{c} \text{Oracle} \\ \text{} \end{array}} &\gate{XH}\gategroup[4,steps=3,style={dashed,rounded corners, inner xsep=0.2pt},background]{{Diffusion Operator}} & \ctrl{1} & \gate{HX}&\ \ldots\ \qw &\gate[wires=4][1cm]{\begin{array}{c} \text{Oracle} \\ \text{} \end{array}} & \gate{XH}\gategroup[4,steps=3,style={dashed,rounded corners, inner xsep=0.2pt},background]{{Diffusion Operator}} & \ctrl{1} & \gate{HX}&\meter{}\\
\lstick{$q_1$}&\qw&\gate{XH} &\ctrl{1}&\gate{HX} &\ \ldots\ \qw&\qw&\gate{XH} &\ctrl{1}&\gate{HX} &\meter{}\\
\lstick{$q_2$}&\qw&\gate{XH} &\ctrl{1}&\gate{HX}&\ \ldots\ \qw&\qw&\gate{XH} &\ctrl{1}&\gate{HX}&\meter{}\\
\lstick{$q_3$}&\qw&\gate{Z} &\targ{}&\gate{Z}&\ \ldots\ \qw&\qw&\gate{Z} &\targ{}&\gate{Z} &\meter{}\qw
\end{quantikz}
\end{adjustbox}
\caption{Quantum circuit of 4-qubit Grover algorithm.}
\label{fig:naivegrovercircuit}
\end{figure}
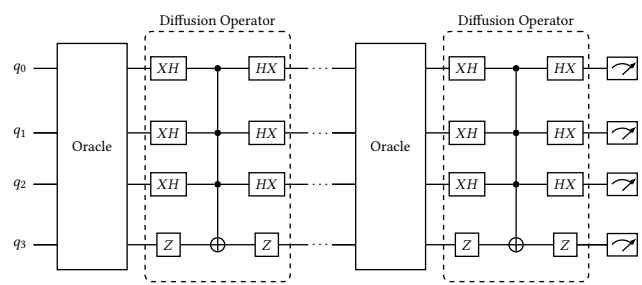
\begin{figure*}[ht]
\centering 
    \begin{subfigure}[t]{0.32\textwidth}
        \includegraphics[width=\columnwidth]{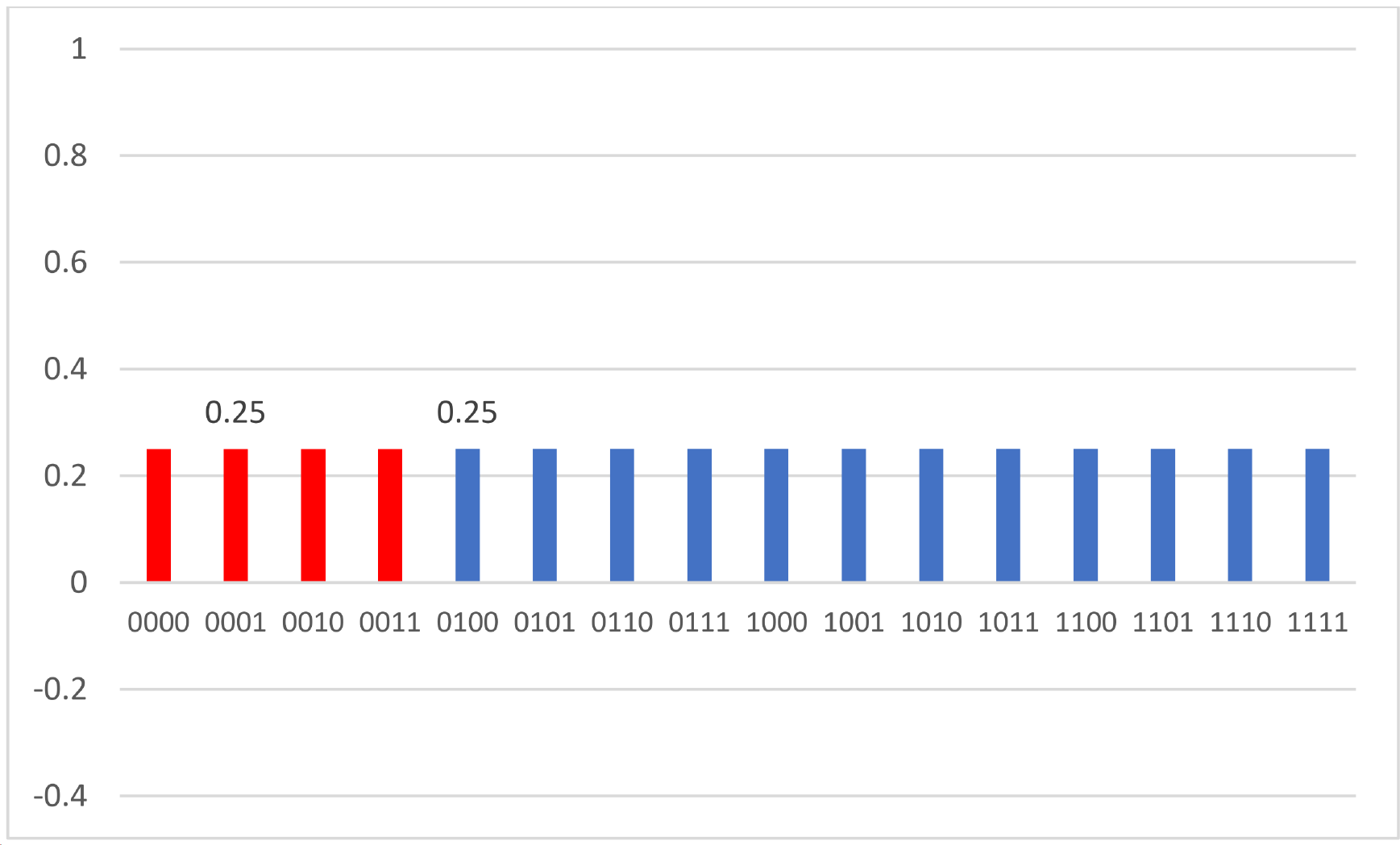}
        \caption{The initial quantum state is an equally distributed superposition state of all items.}
    \end{subfigure}\hfill
    \begin{subfigure}[t]{0.32\textwidth}
        \includegraphics[width=\columnwidth]{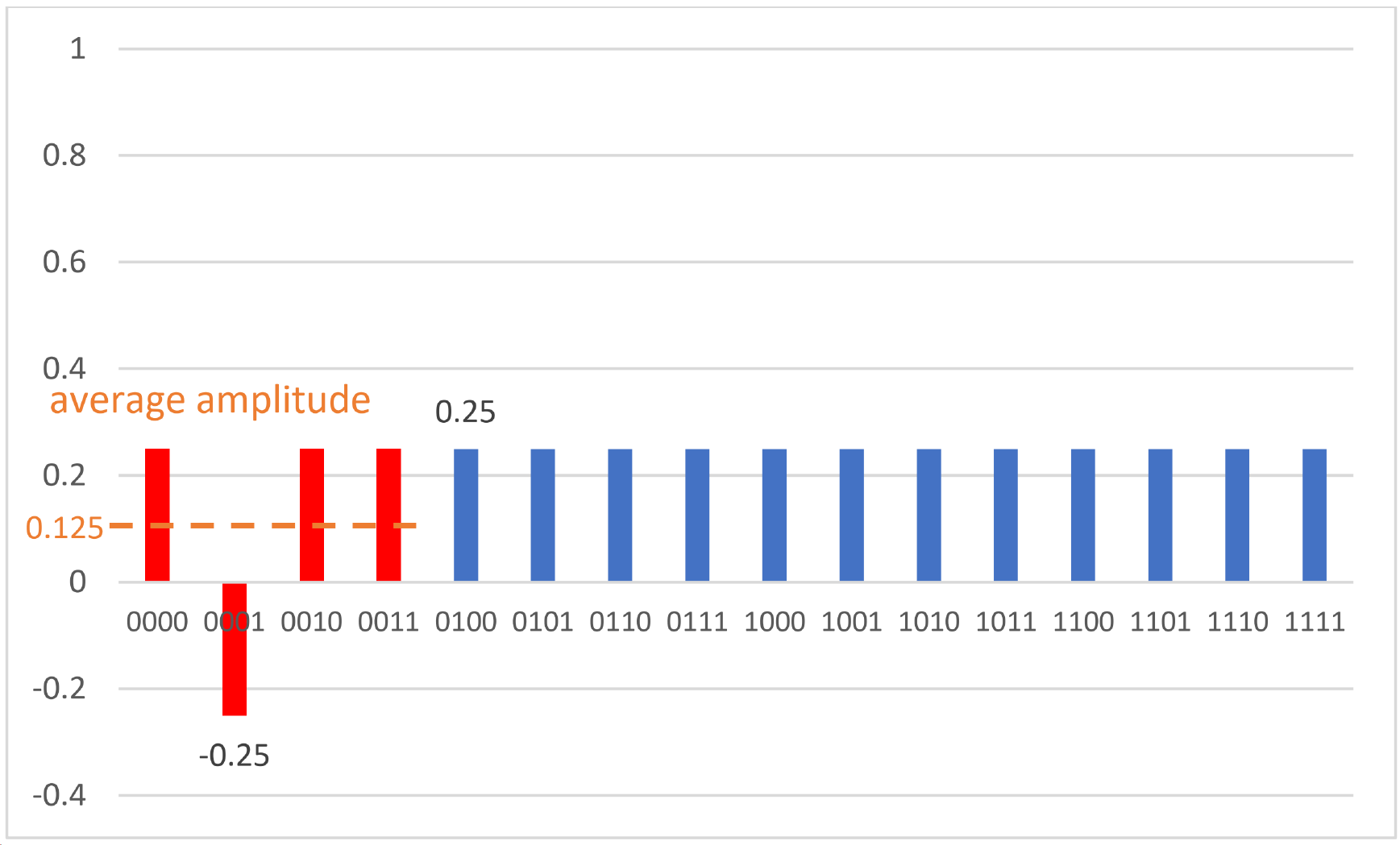}
        \caption{The oracle flips the probability amplitude of the target item.}
    \end{subfigure}\hfill
        \begin{subfigure}[t]{0.32\textwidth}
        \includegraphics[width=\columnwidth]{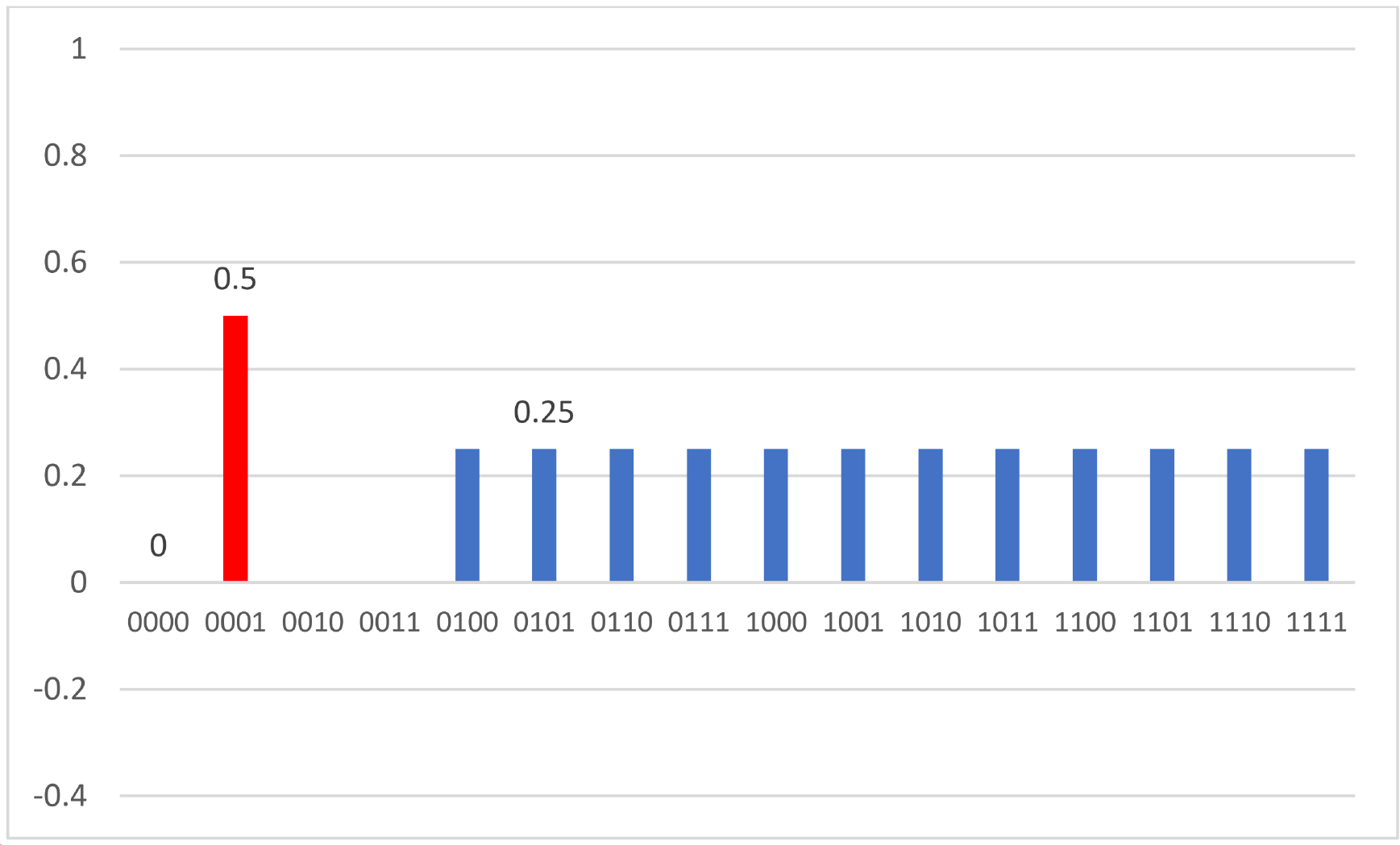}
        \caption{The local diffusion operator inverts the amplitudes about average amplitude in the block.}
        \label{fig:partial3}
    \end{subfigure}\hfill
    \begin{subfigure}[t]{0.4\textwidth}
    \centering
        \includegraphics[width=0.8\columnwidth]{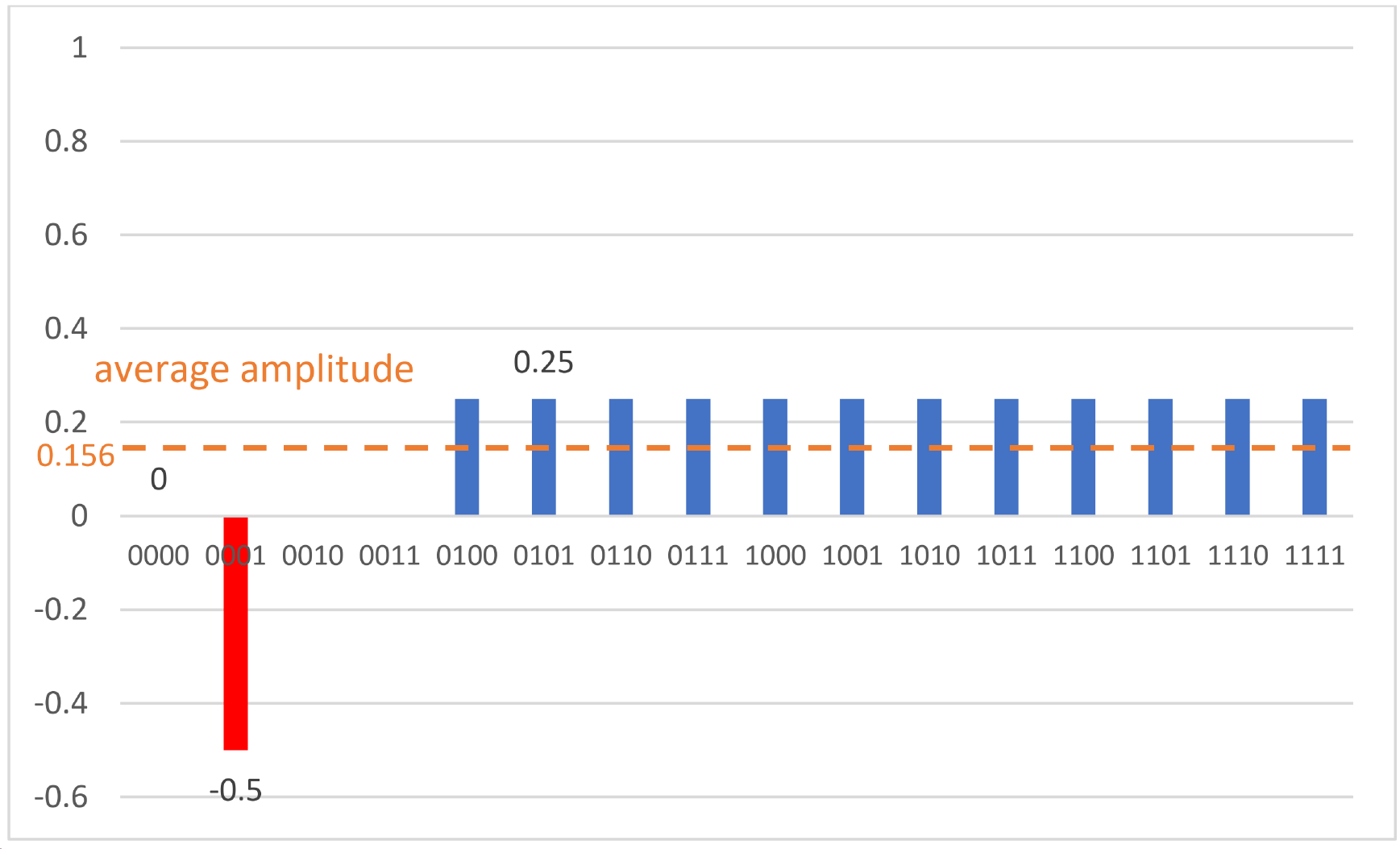}
        \caption{The oracle flips the probability amplitude of the target item.}
    \end{subfigure}\hspace{0.1\textwidth}
        \begin{subfigure}[t]{0.4\textwidth}
        \centering
        \includegraphics[width=0.8\columnwidth]{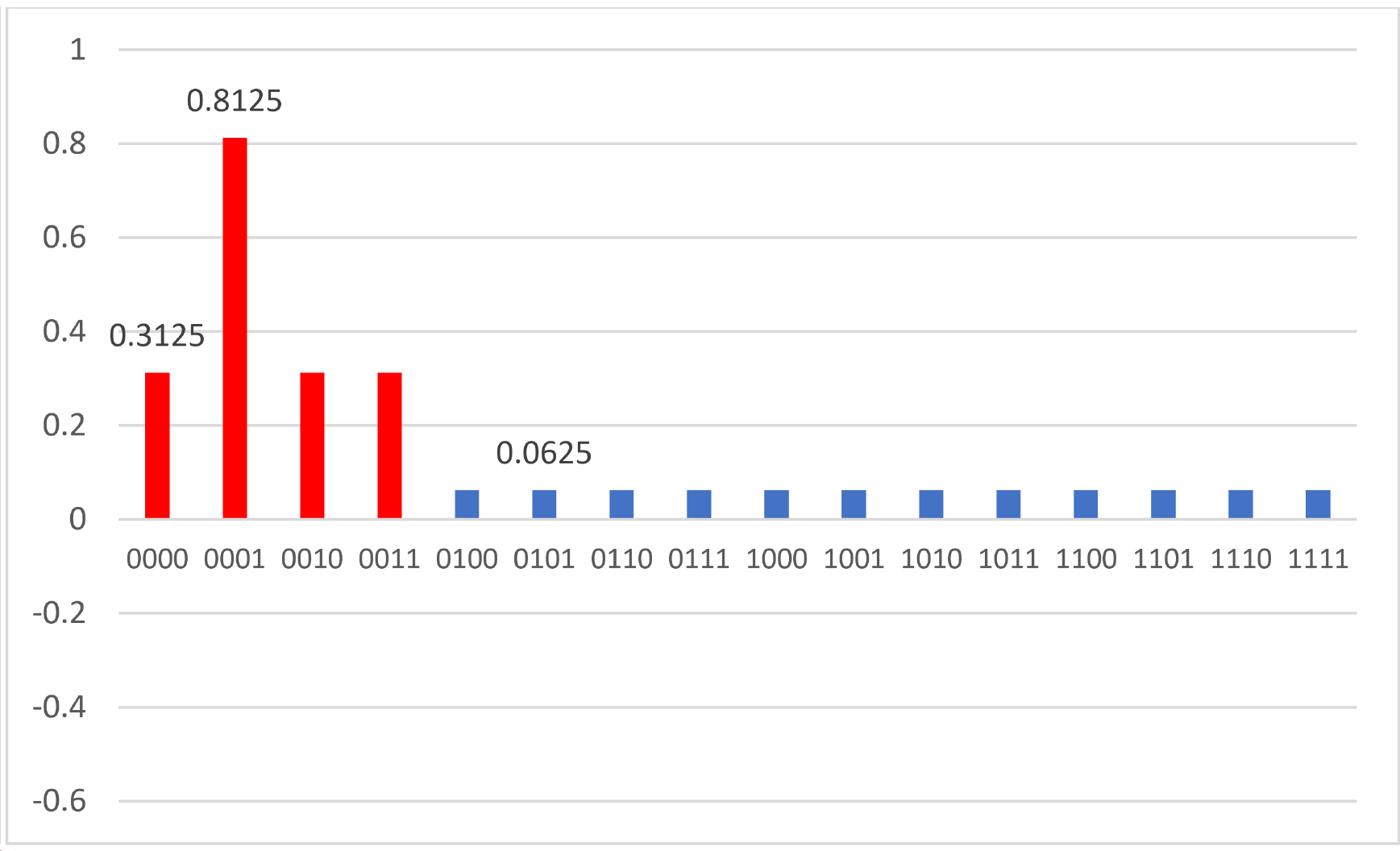}
        \caption{The global diffusion operator inverts the amplitudes about average amplitude of all states, which diminishes the states in non-target blocks}
    \end{subfigure}\hfill
\caption{The first two iterations of the partial Grover's algorithm}
\label{fig:partial_grover_illu}
\end{figure*}

The Grover's operator $G=DO$. In each iteration, we apply the oracle followed by the diffusion operator. After the oracle, the correct solutions will have negative probability amplitudes. Then, we apply the diffusion operator, and the amplitudes will invert about average amplitude. When the average amplitude is greater than $0$, the amplitude of the solutions will be amplified. We use the 4-qubit Grover's algorithm as an example. The first iteration of the algorithm is shown in Figure~\ref{fig:naive_grover_illu}. \textbf{(i)} We initialize the quantum state in the equally distributed superposition of $2^4$ states. The $2^4$ states denote all the items in the database. \textbf{(ii)} We apply the oracle, which flips the probability amplitude of the target item. The flip leads to a decrease in the average amplitude. The target item is $\ket{0001}$. \textbf{(iii)} We apply the diffusion operator. Since the average amplitude is greater than 0, the amplitude of the target item is amplified. The algorithm iteratively applies the oracle and the diffusion operator to amplify the probability amplitude of the target item. If the total number of items is  $N$. After $O(\sqrt{N})$ iterations, the probability of target item will approach 1 with error $O(\frac{1}{\sqrt{N}})$. The exact number of iterations is $\left \lfloor\frac{\pi}{4\theta}-1\right \rfloor$. The algorithm can be extended to cases with multiple targets~\cite{brassard2002Gmultitarget} and sure success~\cite{hu2002suresuccess}.

\subsection{Partial Search Algorithm}
Partial search algorithm, aka partial Grover's algorithm, trades the accuracy of the solution for a higher speed. The partial search algorithm has several variances~\cite{grover2005partial1,korepin2006partial2}. Here we introduce the GRK partial search algorithm, which has been proved to have optimal sequence~\cite{korepin2006partial3}. Similar to Grover's algorithm, the partial search can be extended to multiple targets~\cite{zhang2018partialmultiple} and sure success~\cite{choi2006partialsure}.

The partial search algorithm is an approximate search which finds a target block. A database of $N$ items is divided into $K$ blocks, each block consists of $b$ items: $N = Kb$. We assume both $N$ and $b$ are powers of 2: $N = 2^n$, $b = 2^m$, $k = 2^{n-m}$. Suppose the n-qubit address of the target state is divided into $\ket{t_n} = \ket{t_m}\ket{t_{n-m}}$. Instead of finding the target item, the partial search algorithm finds the target block. In other words, the partial search algorithm finds the items with the last $n-m$ qubits in correct state $\ket{t_{n-m}}$. It does not guarantee the correct solution but requires less query. The oracle complexity of partial search is $O(\sqrt{N}-\sqrt{b})$~\cite{korepin2005grovereigen}. When selecting larger blocks, it requires less number of oracles but the accuracy also gets lower.

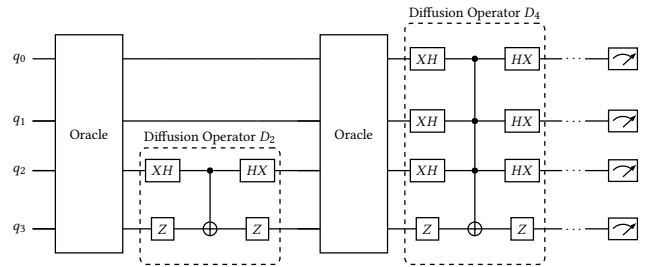
\begin{figure}[ht]
\centering
\begin{adjustbox}{width=\linewidth}
\begin{quantikz}
\lstick{$q_0$} & \gate[wires=4]{\begin{array}{c} \text{Oracle} \\ \text{} \end{array}} &\qw & \qw& \qw& \qw &\gate[wires=4][1cm]{\begin{array}{c} \text{Oracle} \\ \text{} \end{array}} & \gate{XH}\gategroup[4,steps=3,style={dashed,rounded corners, inner xsep=0.2pt},background]{{Diffusion Operator $D_4$}} & \ctrl{1} & \gate{HX}&\ \ldots\ \qw &\meter{}\\
\lstick{$q_1$}&\qw&\qw &\qw&\qw & \qw&\qw&\gate{XH} &\ctrl{1}&\gate{HX} &\ \ldots\ \qw & \meter{}\\
\lstick{$q_2$}&\qw&\gate{XH}\gategroup[2,steps=3,style={dashed,rounded corners, inner xsep=0.2pt},background]{{Diffusion Operator $D_2$}} &\ctrl{1}&\gate{HX}& \qw&\qw&\gate{XH} &\ctrl{1}&\gate{HX}&\ \ldots\ \qw &\meter{}\\
\lstick{$q_3$}&\qw&\gate{Z} &\targ{}&\gate{Z}& \qw&\qw&\gate{Z} &\targ{}&\gate{Z} &\ \ldots\ \qw & \meter{}
\end{quantikz}
\end{adjustbox}
\caption{Quantum circuit of 4-qubit partial search algorithm.}
\label{fig:partialgrovercircuit}
\end{figure}

\begin{figure*}[ht]
\ContinuedFloat
\centering 
    \begin{subfigure}[t]{0.24\textwidth}
        \includegraphics[width=\columnwidth]{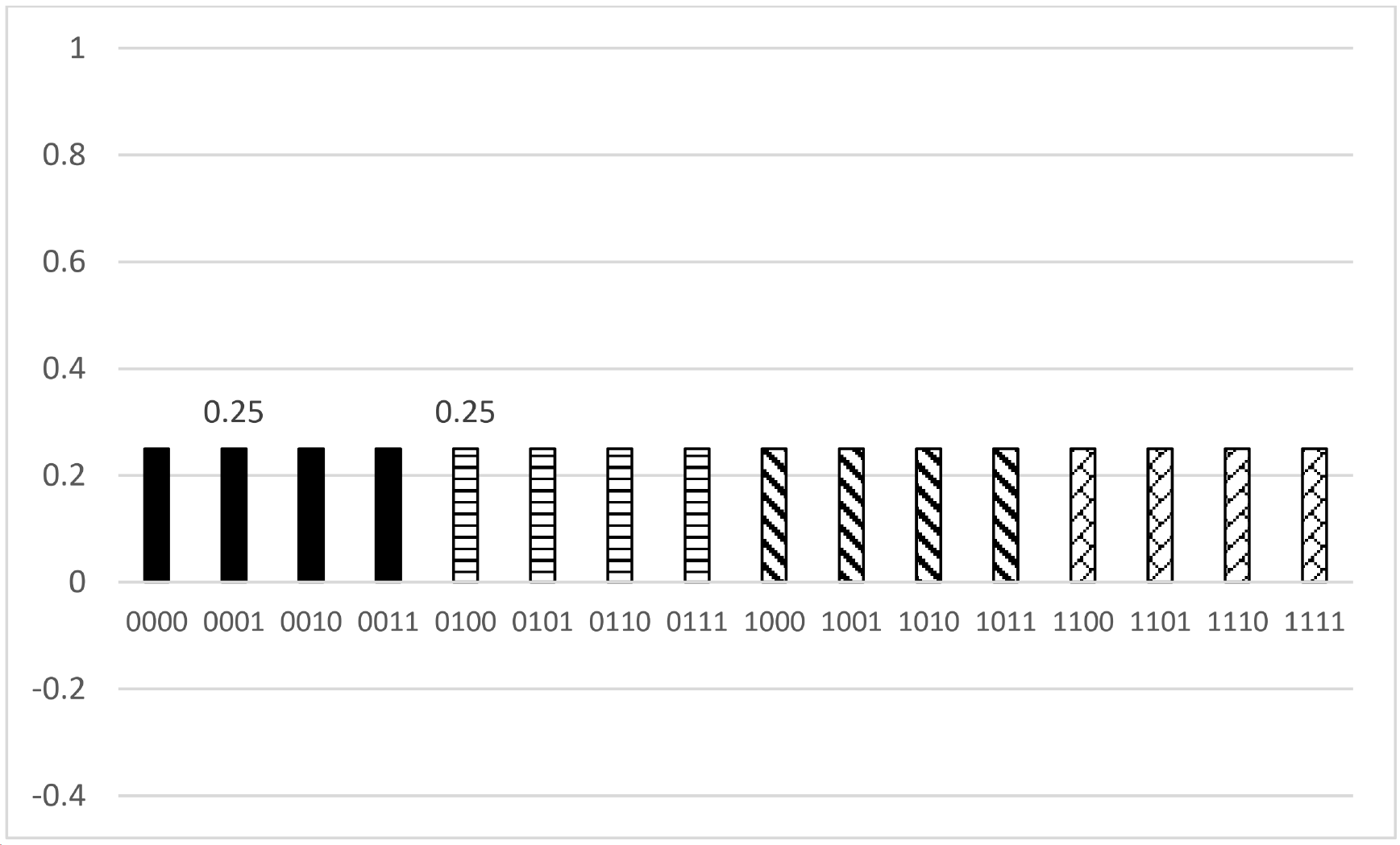}
        \caption{The initial quantum state is an equally distributed superposition state of all items.}
    \end{subfigure}\hfill
    \begin{subfigure}[t]{0.24\textwidth}
        \includegraphics[width=\columnwidth]{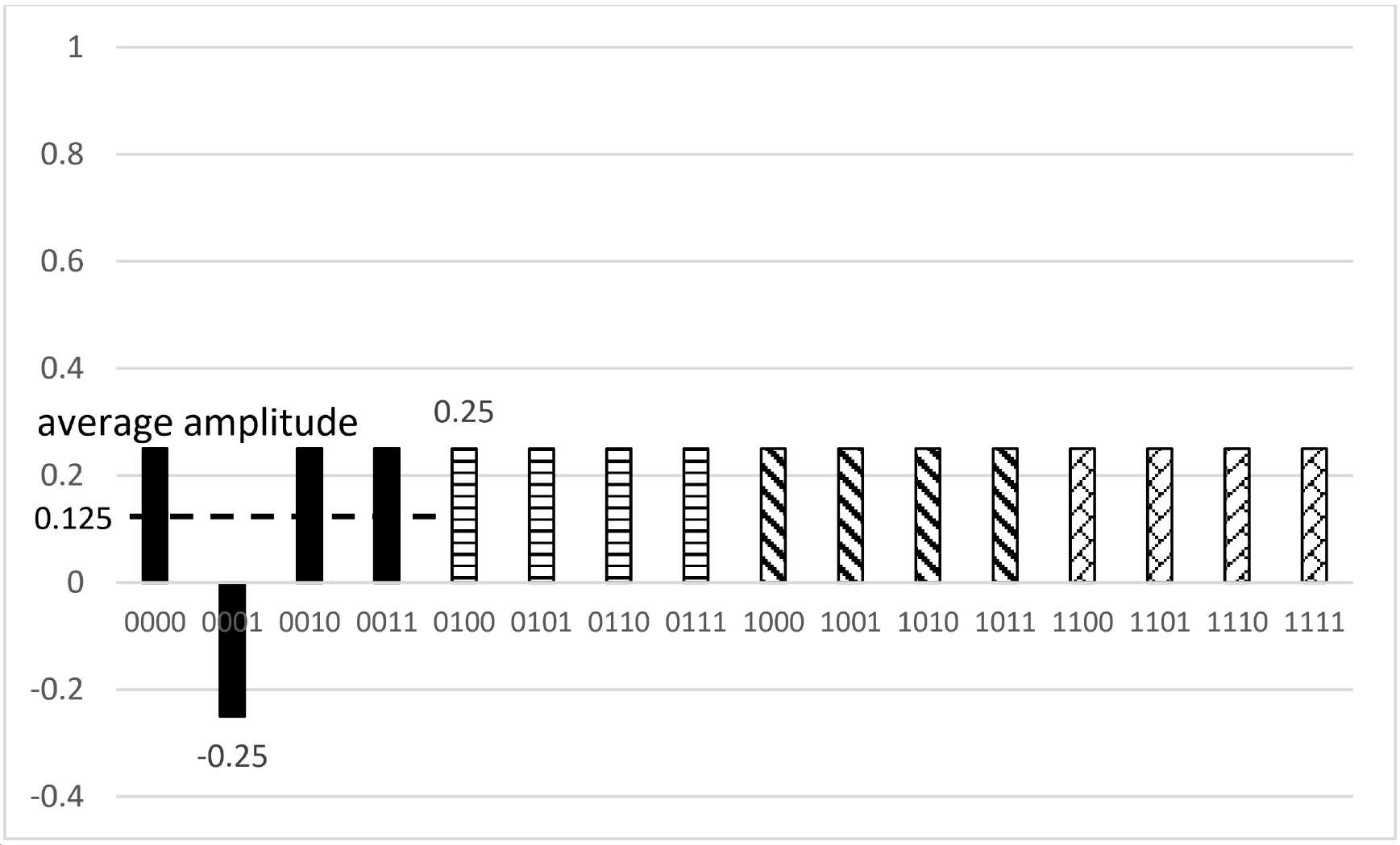}
        \caption{The oracle flips the probability amplitude of the target item.}
        \label{fig:efficient_grover_illu2}
    \end{subfigure}\hfill
        \begin{subfigure}[t]{0.24\textwidth}
        \includegraphics[width=\columnwidth]{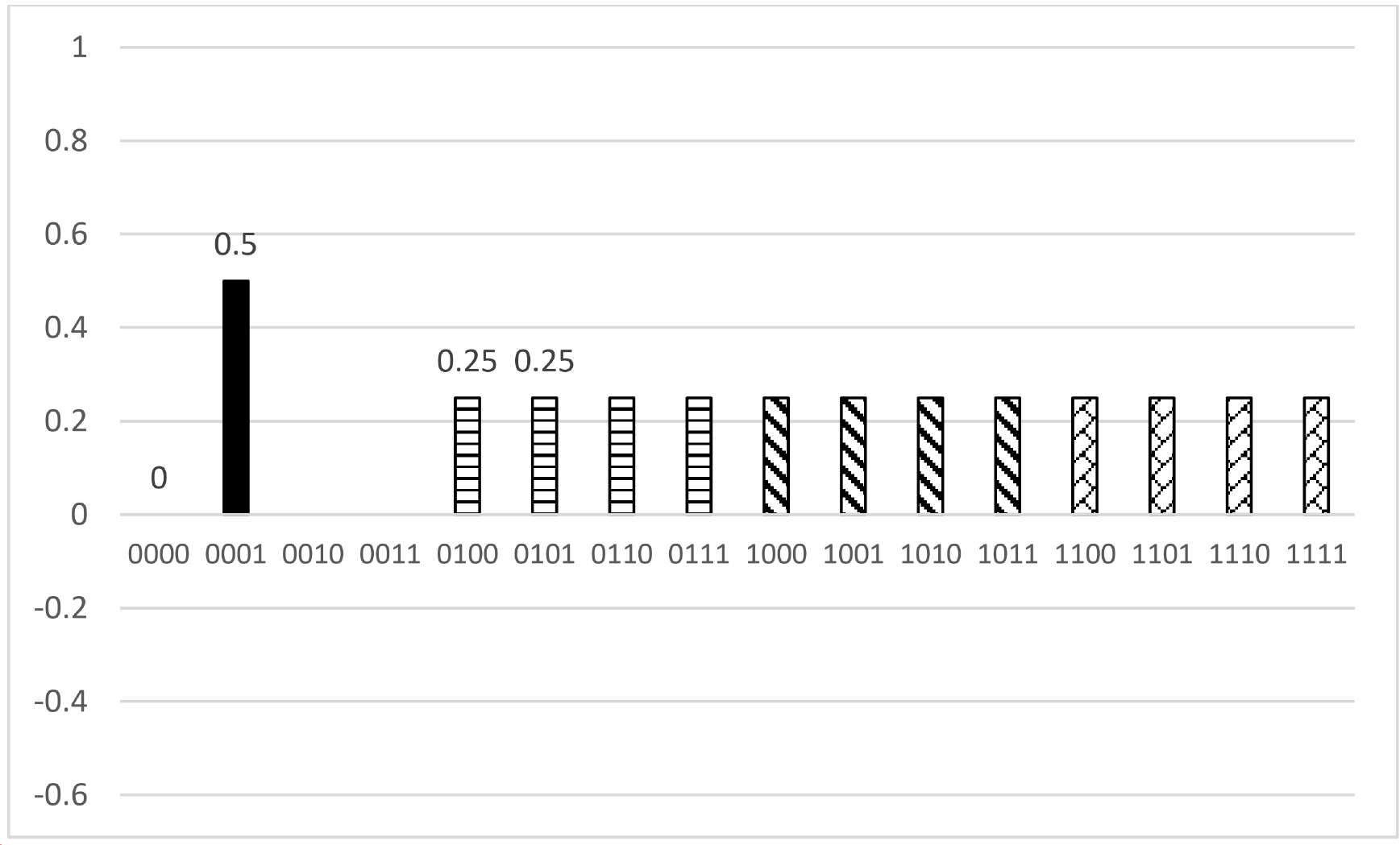}
        \caption{The local diffusion operator $D_2$ inverts the amplitudes about average amplitude in the same target blocks.}
        \label{fig:efficient_grover_illu3}
    \end{subfigure}\hfill
    \begin{subfigure}[t]{0.24\textwidth}
    \centering
        \includegraphics[width=\columnwidth]{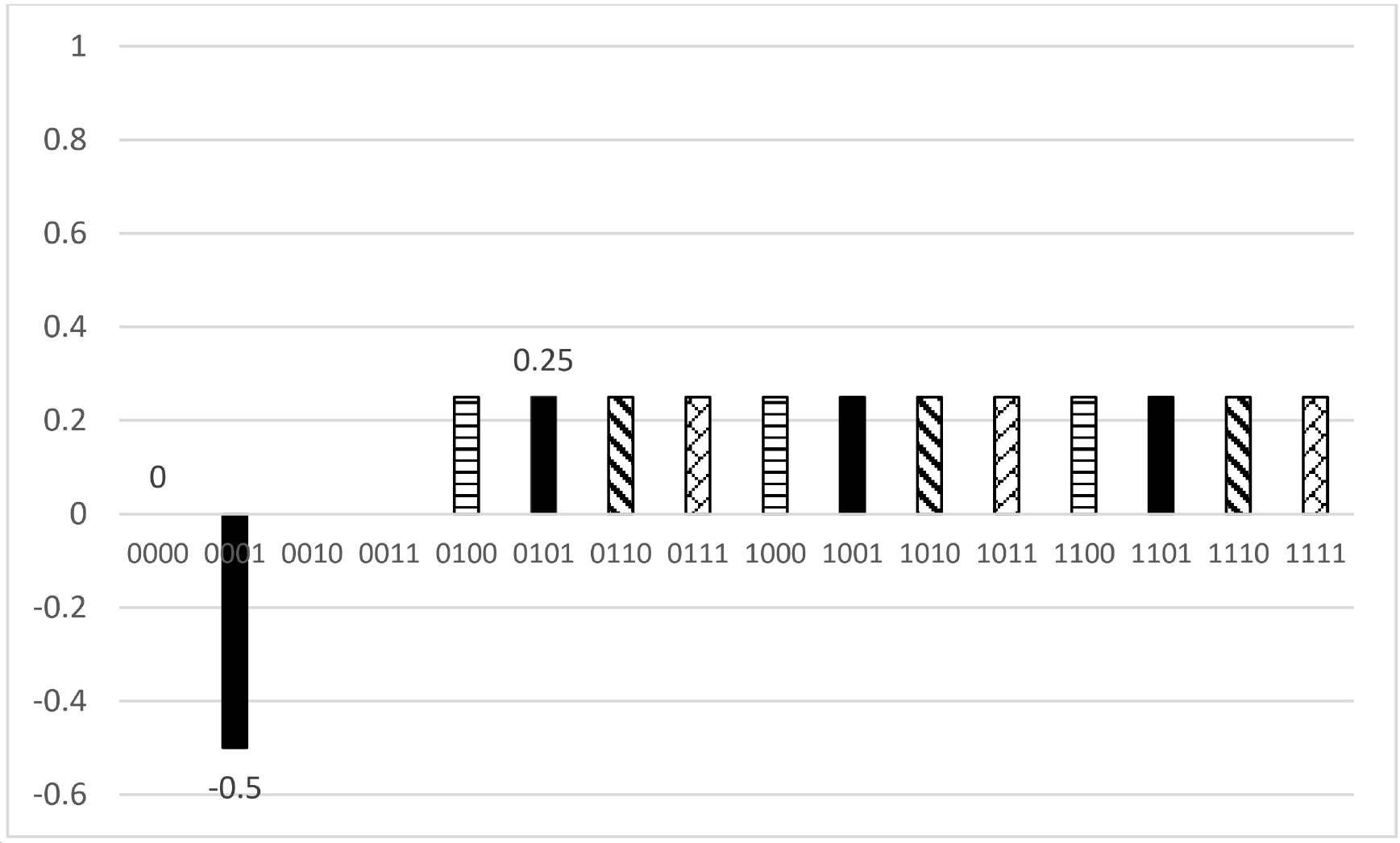}
        \caption{The oracle flips the probability amplitude of the target item.}
        \label{fig:efficient_grover_illu4}
    \end{subfigure}
        \begin{subfigure}[t]{0.32\textwidth}
        \centering
        \includegraphics[width=\columnwidth]{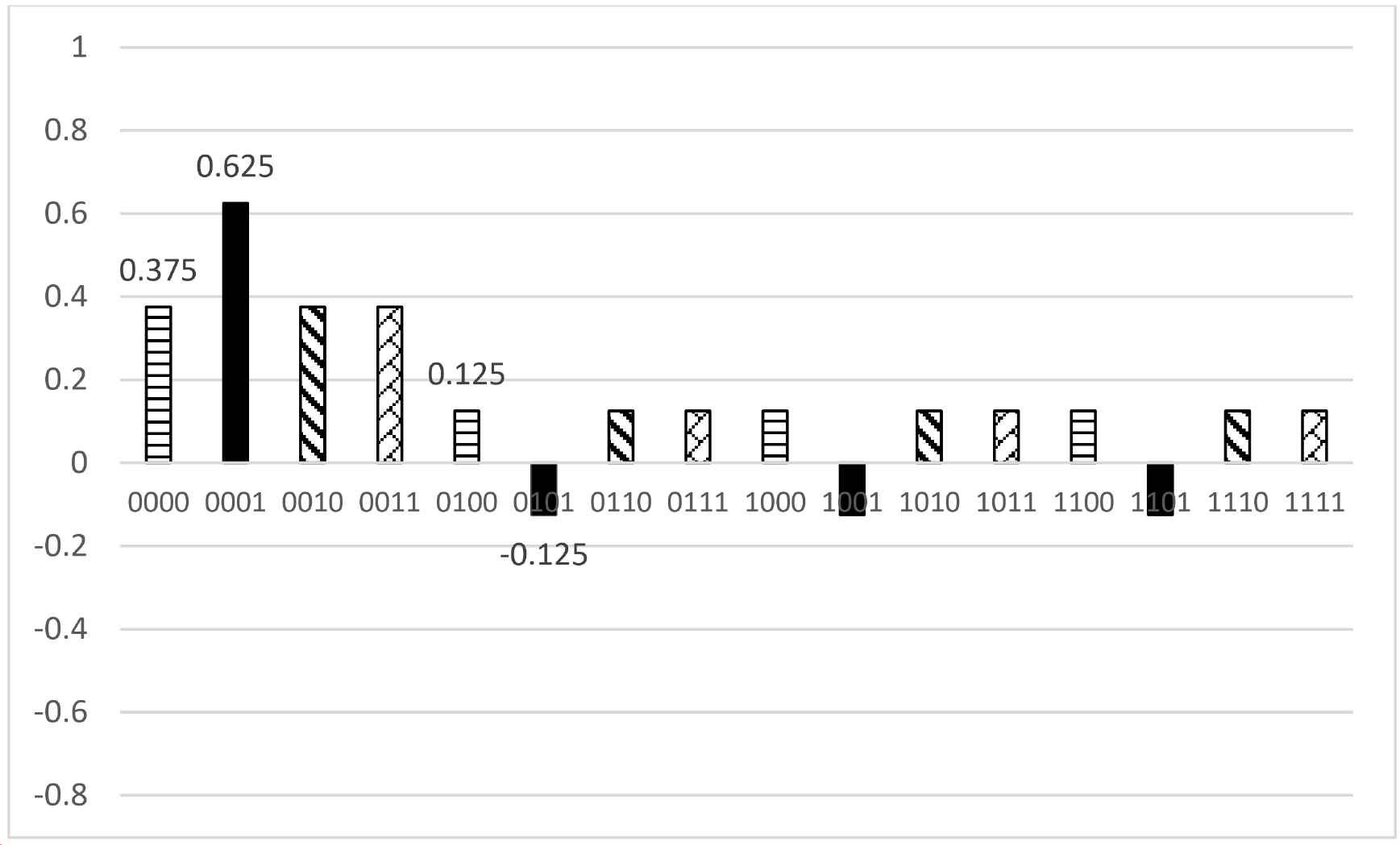}
        \caption{The local diffusion operator $D_2'$ inverts the amplitudes about average amplitude in the new target blocks.}
    \end{subfigure}\hfill
        \begin{subfigure}[t]{0.32\textwidth}
        \centering
        \includegraphics[width=\columnwidth]{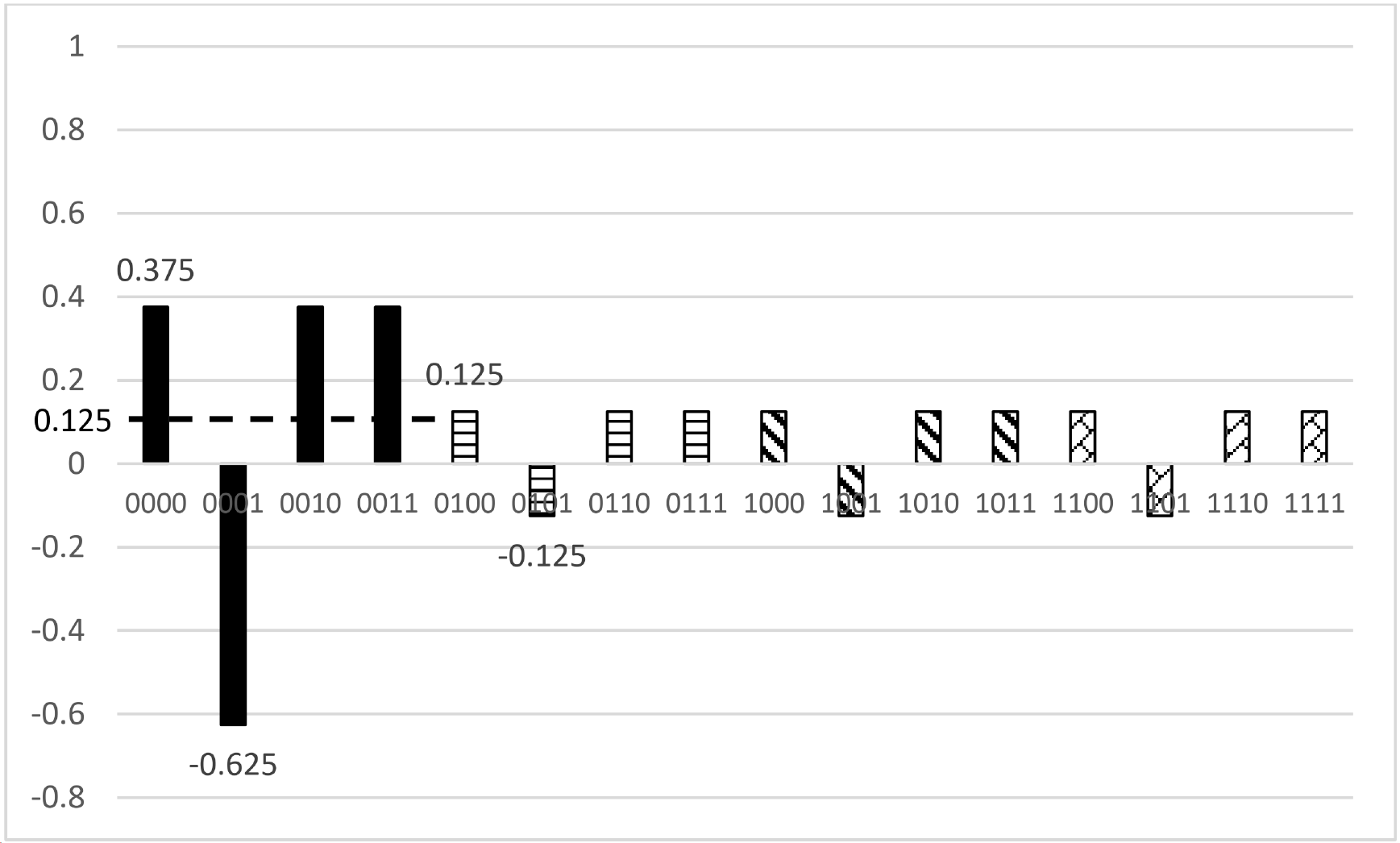}
        \caption{The oracle flips the probability amplitude of the target item.}
    \end{subfigure}\hfill
        \begin{subfigure}[t]{0.32\textwidth}
        \centering
        \includegraphics[width=\columnwidth]{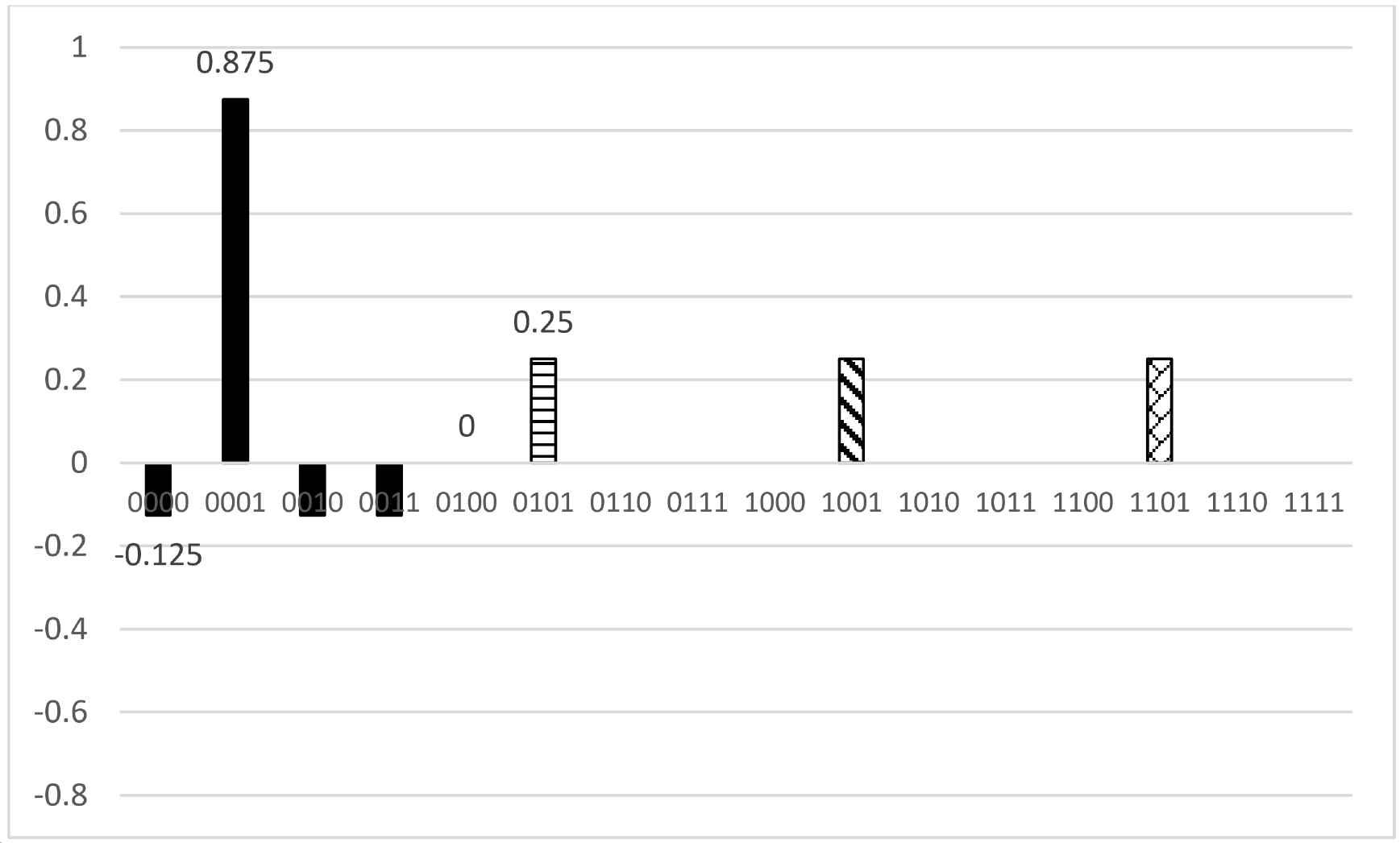}
        \caption{The local diffusion operator $D_2$ inverts the amplitudes about average amplitude in the target blocks.}
    \end{subfigure}\hfill
    
\caption{Amplitude amplification after each operation of our hardware efficient quantum search algorithm, $D_2OD_2'OD_2O$ upon the initial state. The target state is $\ket{0001}$.}
\label{fig:efficient_grover_illu}
\end{figure*}

The partial search algorithm consists of global and local iterations with two different diffusion operators: local diffusion operator $D_m$ and global diffusion operator $D_n$. Figure~\ref{fig:partialgrovercircuit} shows the circuit of local and global iterations for a 4-qubit partial search algorithm with block size $b=2^2$. The probability amplitudes of the first two iterations are illustrated in Figure~\ref{fig:partial_grover_illu}. We mark the states in the target block as red. \textbf{(i)} The qubits are prepared in the equally distributed superposition state. \textbf{(ii)}  The oracle flips the probability amplitude of the target item. The target item is $\ket{0001}$. \textbf{(iii)} The local diffusion operator $D_2$ is applied to the state. The local operator $D_2$ inverts the amplitudes about average amplitude. Here, the average amplitude is the average of the amplitudes within the same block. Note, the other states in the same block (for example $\ket{0100}$,$\ket{0101}$,$\ket{0110}$,$\ket{0111}$) will also invert about their average amplitude. However, since they are equally distributed, the amplitudes will not change. As shown in Figure~\ref{fig:partial3}, the local diffusion operator only changes the states in the target block. \textbf{(iv)} The oracle flips the amplitude of the target item. \textbf{(v)} the global diffusion operator $D_4$ inverts the amplitudes about the average amplitude of all states, which diminishes the states in non-target blocks.
\subsection{Related Work}
With the recent development of quantum hardware, more studies focused on the resource estimation, i.e., width, depth, for Grover's algorithm~\cite{kim2018spacecomplexitygrover,jaques2020groveroracleimplement}. Some studies proposed diffusion operators with depth one. For example, Kato~\cite{kato2005groverdepthone} proposed a Grover-algorithm-like operator using only single-qubit gates. Inspired by the quantum approximate optimization algorithm(QAOA), Jiang et al.~\cite{jiang2017depthone} proposed diffusion operators with depth $O(1)$. However, these algorithms have 0.5 maximal successful probability, and the expected depth is not as efficient as the original Grover's algorithm. Zhang et al.~\cite{zhang2020partialgroverdepth} proposed depth optimization for the partial search algorithm. Wang et al.~\cite{wang2020grovernisq} compared different forms of Grover's algorithm with noise simulation.

\section{Hardware Efficient Search Algorithm}
\label{sec:hardware_efficient_search}
\subsection{Amplitude Amplification with Local Operators}
After reviewing Grover's search algorithm and the partial search algorithm, one natural question is can we perform the quantum search with only local diffusion operators? The answer is 'yes'. Note that this paper focuses on optimizing the diffusion operator. Oracles representing the database functions are usually viewed as black-box functions, so we didn't optimize the oracle. Suppose the n-qubit address of the target state is divided into $\ket{t_n} = \ket{t_m}\ket{t_{n-m}}$. Performing the local Grover search operator $G_m = D_mO$ for the first $m$ qubits will amplify the probabilities of the items with the last $n-m$ qubits in the correct state. Performing the local Grover search operator $G_{n-m}' = D_{n-m}'O$ will amplify the probabilities of the items with first $m$ qubits in the correct state. With these two operators working collectively, we can amplify the probability of the target item. This way, a k-step search operation can be expressed $G = (G_mG_{n-m}')^k = (D_mOD_{n-m}'O)^k$. Note that our algorithm is not the same as directly replacing the global diffusion operator $D_n$ with the local operators $D_m$ and $D_{n-m}'$, which would lead to the search operation $G = (D_mD_{n-m}'O)^k$. 

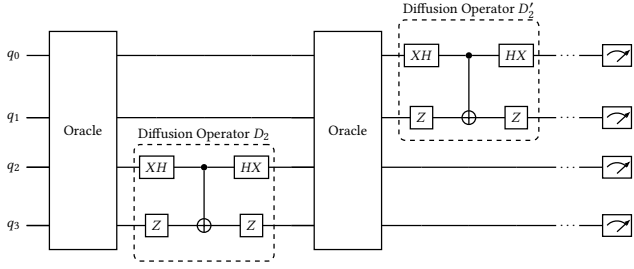
\begin{figure}[ht]
\centering
\begin{adjustbox}{width=\linewidth}
\begin{quantikz}
\lstick{$q_0$} & \gate[wires=4]{\begin{array}{c} \text{Oracle} \\ \text{} \end{array}} &\qw & \qw& \qw& \qw &\gate[wires=4][1cm]{\begin{array}{c} \text{Oracle} \\ \text{} \end{array}} & \gate{XH}\gategroup[2,steps=3,style={dashed,rounded corners, inner xsep=0.2pt},background]{{Diffusion Operator $D'_2$}} & \ctrl{1} & \gate{HX}&\ \ldots\ \qw &\meter{}\\
\lstick{$q_1$}&\qw&\qw &\qw&\qw & \qw&\qw&\gate{Z} &\targ{}&\gate{Z} &\ \ldots\ \qw & \meter{}\\
\lstick{$q_2$}&\qw&\gate{XH}\gategroup[2,steps=3,style={dashed,rounded corners, inner xsep=0.2pt},background]{{Diffusion Operator $D_2$}} &\ctrl{1}&\gate{HX}& \qw&\qw&\qw &\qw&\qw&\ \ldots\ \qw &\meter{}\\
\lstick{$q_3$}&\qw&\gate{Z} &\targ{}&\gate{Z}& \qw&\qw&\qw &\qw&\qw &\ \ldots\ \qw & \meter{}
\end{quantikz}
\end{adjustbox}
\caption{The quantum circuit of first two iterations of 4-qubit hardware efficient quantum search algorithm.}
\label{fig:efficientgrovercircuit}
\end{figure}

The quantum circuit of the first two iterations of our hardware efficient quantum search algorithm is shown in Figure~\ref{fig:efficientgrovercircuit}. The search algorithm consists of two local Grover searches with round-robin scheduling. These two local searches $G_m$ and $G_{n-m}'$ operate on different qubits, and the collaboration of two local searches covers all the items. 

The probability amplitudes of the first three iterations are illustrated in Figure~\ref{fig:efficient_grover_illu}. \textbf{(i)} The qubits are initialized in the equally distributed superposition state. The amplitudes with the same pattern are in the same local search block. \textbf{(ii)} The oracle flips the probability amplitude of the target item. The target item is $\ket{0001}$. \textbf{(iii)} The local diffusion operator $D_2$ operates on the last two qubits. Therefore, states with the first two qubits in the same state belong to the same target block. For example, $\ket{0000},\ket{0001},\ket{0010}, \ket{0011}$ belong to the same target block. The amplitude of the states will invert about the average amplitude of the states in the same block. \textbf{(iv)} The oracle flips the probability amplitude of the target item. We changed the pattern to represent the new target blocks. \textbf{(v)} The local diffusion operator $D_2'$ operates on the first two qubits. Therefore, states with the last two qubits in the same state belong to the same target block. For example, $\ket{0001},\ket{0101},\ket{1001}, \ket{1101}$ belongs to the same target block. The amplitude of the states will invert about the average amplitude of the states in the same block. \textbf{(vi)} The oracle flips the probability amplitude of the target item. The patterns are changed to represent the new target blocks. \textbf{(vii)} The local diffusion operator $D_2$ for the first two qubits further amplifies the amplitude of the target item.

In general, for problem with n-qubit address, we can divide the target state into: $\ket{t} = \ket{t_m}\ket{n-m}$. We iteratively apply the local Grover operator $G_m$ and $G_{n-m}'$ to the quantum state. The states can be classified into four sets: $S(t)$: The set of the target states, $S(nt_m)$: The of the non-target states that belongs to the target block in the first search. In other words, the states with leading $n-m$ qubits same as the target state. For example, state $\ket{0000}$, $\ket{0010}$, and $\ket{0011}$ in Figure~\ref{fig:efficient_grover_illu2}. $S(nt_{n-m})$: The set of the non-target states that belongs to the target block in the second search. In other words, the states with the last $m$ qubits same as the target state. For example, state $\ket{0101}$, $\ket{1001}$, and $\ket{1101}$ in Figure~\ref{fig:efficient_grover_illu4}. $S(u)$: The set of the states not in either target blocks. 

\begin{figure}[ht]
\centering 
    \begin{subfigure}[bt]{0.48\linewidth}
        \includegraphics[width=\columnwidth, height = 60mm]{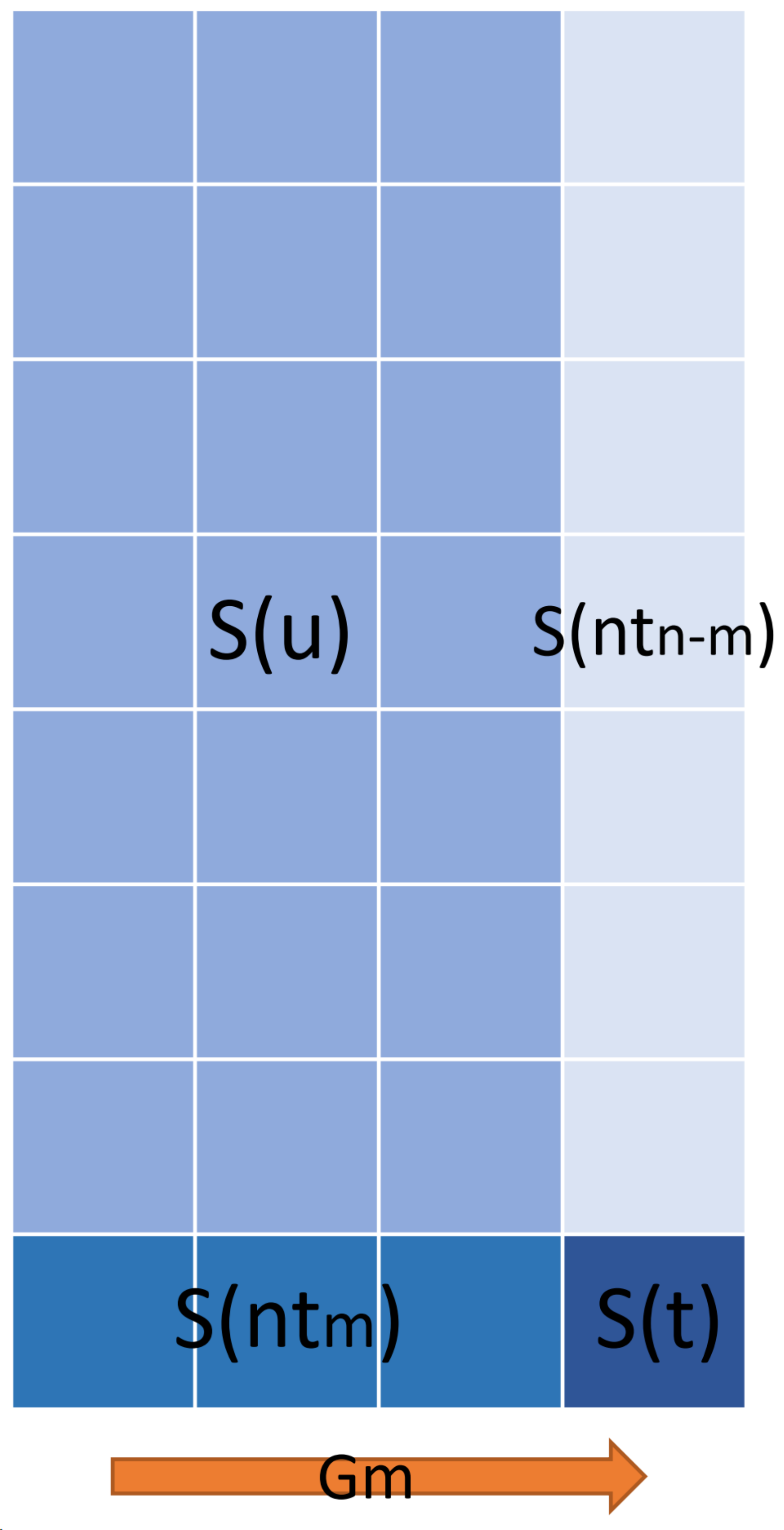}
        \caption{The $G_{m}$ operator amplifies the amplitude of states in $S(t)$ and $S(nt_{n-m})$}
        \label{subfig:operatorm}
    \end{subfigure}\hfill
    \begin{subfigure}[bt]{0.48\linewidth}
    \vspace*{-4mm}
        \includegraphics[width=\columnwidth, height = 57mm]{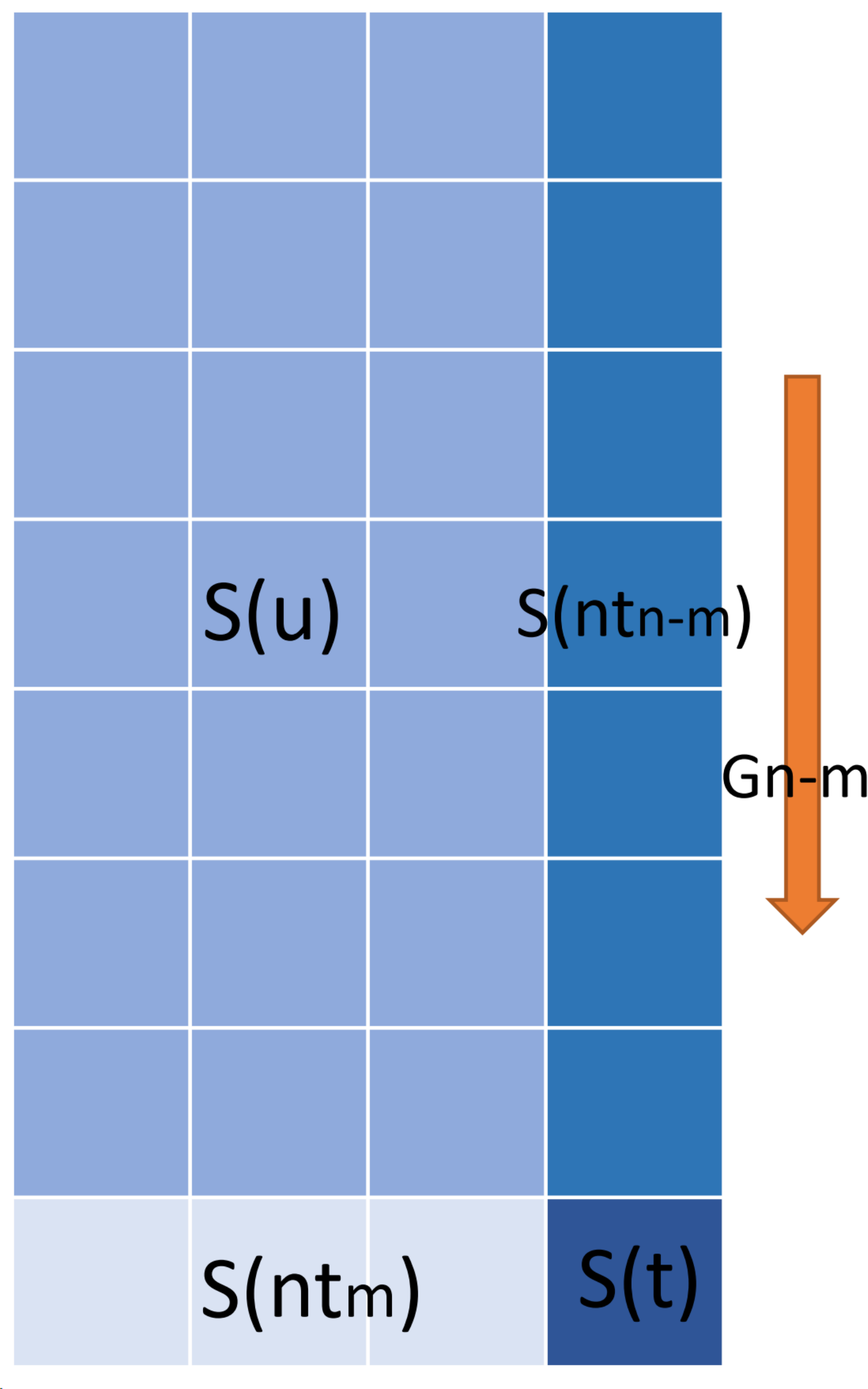}
        \caption{The $G_{n-m}$ operator amplifies the amplitude of states in $S(t)$ and $S(nt_{m})$}
        \label{subfig:operatornm}
    \end{subfigure}\hfill
\caption{The effect of the local search operators. Each grid represents an item. The items with darker color has higher probability amplitude. The arrows show the directions of the probability amplitude convergence.}
\label{fig:effectoflocal}
\end{figure}

Figure~\ref{fig:effectoflocal} demonstrates the effect of the local search operators. Each grid represents an item. Each row has $2^m$ items, and each column has $2^{n-m}$ items. The grids in the same row denote the items in the same block during the first local search $G_m$. The grids in the same column denote the items in the same block during the second local search $G_{n-m}$. As shown in Figure~\ref{subfig:operatorm}, applying the local operator $G_m$ has two effects. First, it performs a local Grover search for the states from $S(t)$ and $S(nt_m)$. It will amplify the amplitudes of the states in $S(t)$ and diminish the amplitudes of the states in $S(nt_m)$. Second, it flips the states in $S(u)$ and the states in $S(nt_{n-m})$ about their average amplitude. It will amplify the amplitudes of the states in $S(nt_{n-m})$. After the $G_m$ operator, the state distribution will be similar to the distribution in Figure~\ref{subfig:operatornm}. Applying the $G_{n-m}$ will amplify the amplitudes of the states in $S(t)$ and $S(nt_{m})$. And the state distribution returns to the distribution in Figure~\ref{subfig:operatorm}. By iteratively applying $G_m$ and $G_{n-m}$, we will amplify the amplitude of state $\ket{t}$. Our algorithm can be viewed as a combination of two local searches with block size $2^m$ and $2^{n-m}$. The oracle complexity is $O(\sqrt{2^m})\times O(\sqrt{2^{n-m}}) = O(\sqrt{2^n})$

 .  
\subsection{Matrix Representation}
In this section, we will derive the matrix representation of our hardware efficient algorithm. Assume we have $n$ qubits. We partition the search into two local searches with $m$, and $n-m$ qubits respectively. When the qubits are in the equally distributed state, the probability amplitude of each state is: $\frac{1}{\sqrt{N}}$, here $N = 2^{n}$. In general, the algorithm consists of a sequence of operators: $G=(G^{k2}_{n-m}G^{k1}_{m})^k$ . Assume there is one solution $\ket{t}$. We define three normalized states to represent the states in set $S(nt_m)$, $S(nt_{n-m})$ and $S(u)$. $\ket{nt_m}$ is the normalized sum of all non-target states in the first target block, $\ket{nt_{n-m}}$ is the normalized sum of all states in the second target block, $\ket{u}$ is the normalized sum of all states that are not in either target blocks:

\begin{flalign}
&\ket{nt_m} = \frac{1}{\sqrt{2^m-1}}\sum_{first\ target,x\neq t}\ket{x}\\
&\ket{nt_{n-m}} =\frac{1}{\sqrt{2^{n-m}-1}}\sum_{second\ target,x\neq t}\ket{x}\\
&\ket{u} =\frac{1}{\sqrt{2^n-2^m-2^{n-m}+1}}\sum_{non-target}\ket{x}
\end{flalign}

The initial equally distributed state $\ket{\psi}$ may be re-expressed as:
\begin{equation} 
\begin{aligned}
\ket{\psi} = \sqrt{\frac{1}{2^n}}\ket{t} + \sqrt{\frac{2^{m}-1}{2^n}}\ket{nt_m} + \sqrt{\frac{2^{n-m}-1}{2^n}}\ket{nt_{n-m}}\\
+ \sqrt{\frac{2^n-2^{m}-2^{n-m} + 1}{2^n}}\ket{u}
\end{aligned}
\end{equation}

Let $sin(\theta) = \sqrt{\frac{1}{2^{m}}}$, $cos(\theta) = \sqrt{\frac{2^m-1}{2^{m}}}$, and $sin(\gamma) = \sqrt{\frac{2^m}{2^n}}$, $cos(\gamma) = \sqrt{\frac{2^n-2^{m}}{2^n}}$.

\begin{equation} \label{eq:7}
\begin{aligned}
\ket{\psi} =sin(\gamma) (sin(\theta)\ket{t} + cos(\theta)\ket{nt_m})\\
+ cos(\gamma)(sin(\theta)\ket{nt_{n-m}} + cos(\theta)\ket{u})
\end{aligned}
\end{equation}







The global diffusion operator $D_{n}$ for the $n$ qubits can be expressed as~\cite{nielsen2002quantumcomputation}:
\begin{equation} 
D_{n} = 2\ket{\psi}\bra{\psi} - I^{\otimes n}
\end{equation}
where $\ket{\psi}$ is the normalized equally distributed superposition state.


The local operator $D_{m}$ for the first $m$ qubits can be expressed as:
\begin{equation} 
D_{m} = (2\ket{\psi}_m\bra{\psi}_m - I^{\otimes m})\otimes I^{\otimes n-m}
\end{equation}
where $\ket{\psi}_{m}$ is the equally distributed superposition state. The local Grover's operator is the combination of the local diffusion operator and the oracle: $G_m=D_mO$.
The four relevant eigenvectors of $G_{m}$ are~\cite{korepin2005grovereigen}:

\begin{equation} 
\begin{aligned}
&G_{m}\ket{\psi_m^\pm} = \lambda_m^\pm\ket{\psi_m^\pm}\\
&G_{m}\ket{\psi_m^0} = \ket{\psi_m^0}\\
&G_{m}\ket{\psi_m^1} = -\ket{\psi_m^1}
\end{aligned}
\end{equation}
here $\ket{\psi_m^0}$ is the normalized sum of the equally distributed states not in the first target block. Since $G_m$ only inverts the state about average amplitude, it has no effect on $\ket{\psi_m^0}$. $\ket{\psi_m^1}$ is the normalized sum of the states not in the first block but with average amplitude equals to 0. Therefore, after applying $G_m$ the amplitude of the individual state will flip. The eigenvectors are:
\begin{equation} 
\begin{aligned}
&\ket{\psi_m^\pm}=\frac{1}{\sqrt{2}}\ket{t}\pm\frac{i}{\sqrt{2}}\ket{nt_m},\lambda_m^\pm=e^{\pm2i\theta}\\
&\ket{\psi_m^0} = sin(\theta)\ket{nt_{n-m}} + cos(\theta)\ket{u}\\
&\ket{\psi_m^1} = cos(\theta)\ket{nt_{n-m}} - sin(\theta)\ket{u}\\
\end{aligned}
\end{equation}


The normalized vectors can be re-expressed with the eigenvectors:

\begin{equation} 
\begin{aligned}
&\ket{t} = \frac{1}{\sqrt{2}}(\ket{\psi_m^+} + \ket{\psi_m^-})\\
&\ket{nt_m} = \frac{-i}{\sqrt{2}}(\ket{\psi_m^+}-\ket{\psi_m^-})\\
&\ket{nt_{n-m}} =sin(\theta)\ket{\psi_m^0} + cos(\theta)\ket{\psi_m^1}\\
&\ket{u} = cos(\theta)\ket{\psi_m^0} - sin(\theta)\ket{\psi_m^1}
\end{aligned}
\end{equation}


The initial state is the equally distributed state $\ket{\psi}$. It can be represented with the eigenvectors:
\begin{equation} 
\begin{aligned}
\ket{\psi} = \frac{isin(\gamma)}{\sqrt{2}}(-e^{i\theta}\ket{\psi_m^+} +
e^{-i\theta}\ket{\psi_m^-}) + cos(\gamma)\ket{\psi_m^0}
\end{aligned}
\end{equation}

After applying the $G_{m}$ operator for $k_1$ times, the state is:

\begin{equation}
\resizebox{.9\linewidth}{!}{$
\begin{aligned}G_{m}^{k_1}\ket{\psi} = sin(\lambda) (sin((2k_1+1)\theta)\ket{t} + cos((2k_1+1)\theta)\ket{nt_m})\\
+ cos(\lambda)(sin(\theta)\ket{nt_{n-m}} + cos(\theta)\ket{u})
\end{aligned}$}
\end{equation}

Then, we apply the local Grover operator $G_{n-m}$ for $k_2$ times to further amplify the probability of the solution state and diminish the probability of the states outside the target block. $G_{n-m}$ operates on the following $n-m$ qubits. Similar to the $G_{m}$ operator, we have four relevant eigenvectors of $G_{n,m}$:

\begin{equation} 
\begin{aligned}
&G_{n-m}\ket{\psi_{n-m}^\pm} = \lambda_{n-m}^\pm\ket{\psi_{n-m}^\pm}\\
&G_{n-m}\ket{\psi_{n-m}^0} = \ket{\psi_{n-m}^0}\\
&G_{n-m}\ket{\psi_{n-m}^1} = -\ket{\psi_{n-m}^1}
\end{aligned}
\end{equation}
eigenvectors and eigenvalues are:
\begin{equation} 
\begin{aligned}
&\ket{\psi_{n-m}^\pm}=\frac{1}{\sqrt{2}}\ket{t}\pm\frac{i}{\sqrt{2}}\ket{nt_{n-m}},\lambda_{n-m}^\pm=e^{\pm2i\gamma}\\
&\ket{\psi_{n-m}^0} = sin(\gamma)\ket{nt_{m}} + cos(\gamma)\ket{u}\\
&\ket{\psi_{n-m}^1} = cos(\gamma)\ket{nt_{m}} - sin(\gamma)\ket{u}\\
\end{aligned}
\end{equation}


The normalized vectors can be re-expressed with the eigenvectors:

\begin{equation} 
\begin{aligned}
&\ket{t} = \frac{1}{\sqrt{2}}(\ket{\psi_{n-m}^+} + \ket{\psi_{n-m}^-})\\
&\ket{nt_m} =sin(\gamma)\ket{\psi_{n-m}^0} + cos(\gamma)\ket{\psi_{n-m}^1}\\
&\ket{nt_{n-m}} = \frac{-i}{\sqrt{2}}(\ket{\psi_{n-m}^+}-\ket{\psi_{n-m}^-})\\
&\ket{u} = cos(\gamma)\ket{\psi_{n-m}^0} - sin(\gamma)\ket{\psi_{n-m}^1}
\end{aligned}
\end{equation}

Two sets of eigenvectors form two basises. We can represent Grover's search operator in the matrix form with respect to the basis of their eigenvectors:
\begin{equation}
\resizebox{0.9\linewidth}{!}{$
G_{m}' = \begin{pmatrix}
e^{2i\theta}& 0& 0&0\\
0 & e^{-2i\theta} &0 &0\\
0 & 0& 1 & 0\\
0 & 0& 0 & -1
\end{pmatrix},G_{n-m}' = \begin{pmatrix}
e^{2i\gamma}& 0& 0&0\\
0 & e^{-2i\gamma} &0 &0\\
0 & 0& 1 & 0\\
0 & 0& 0 & -1
\end{pmatrix}$}
\end{equation}

\begin{figure*}[t]
    \centering 
    \begin{subfigure}[t]{0.4\textwidth}
    \centering
    \includegraphics[width=\textwidth]{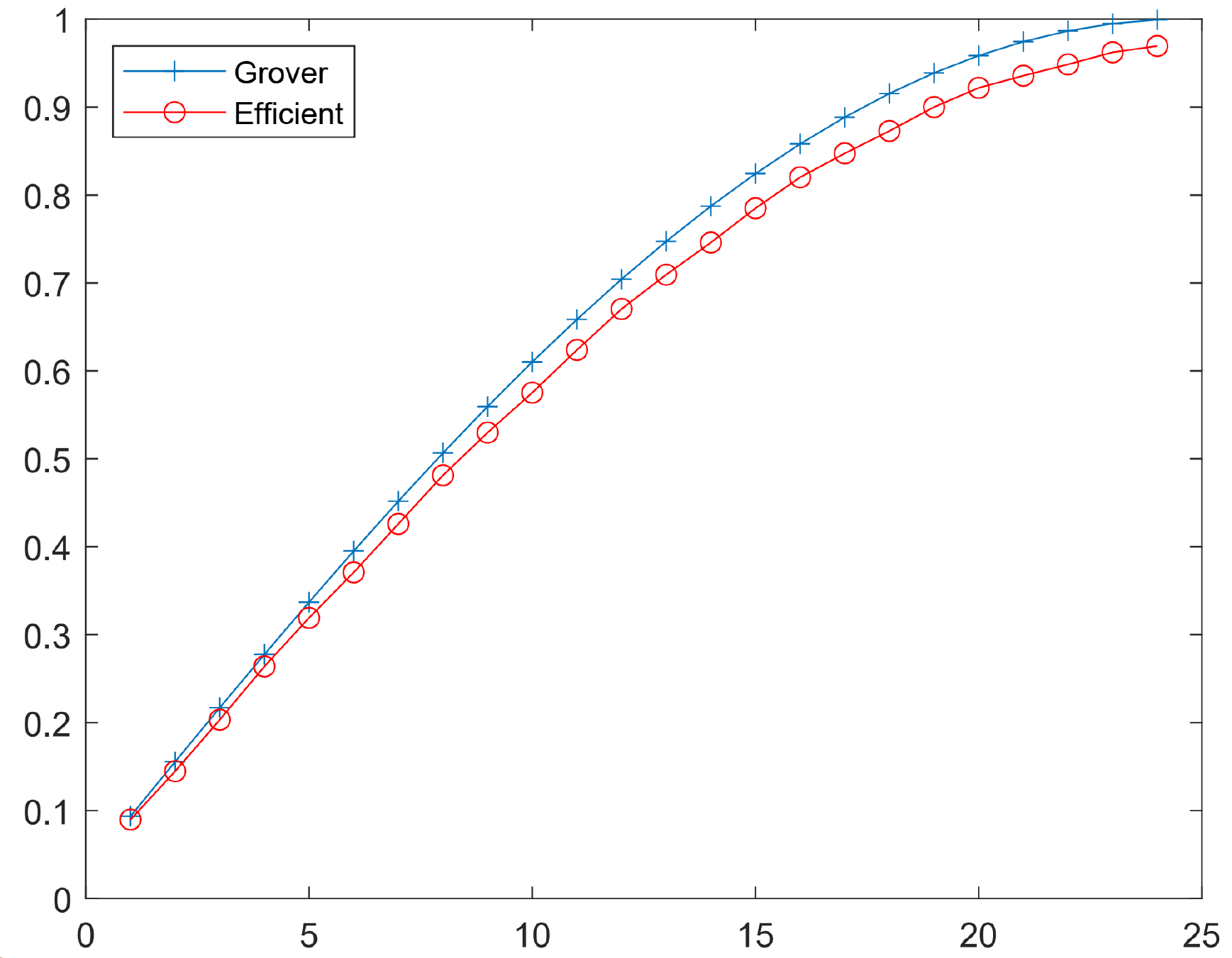}
    \caption{$n=10,m=5,k_1=k_2=1$}
    \label{subfig:grover_5_10_11}
    \end{subfigure}\hspace{2mm}
    \begin{subfigure}[t]{0.4\textwidth}
    \centering
    \includegraphics[width=\columnwidth]{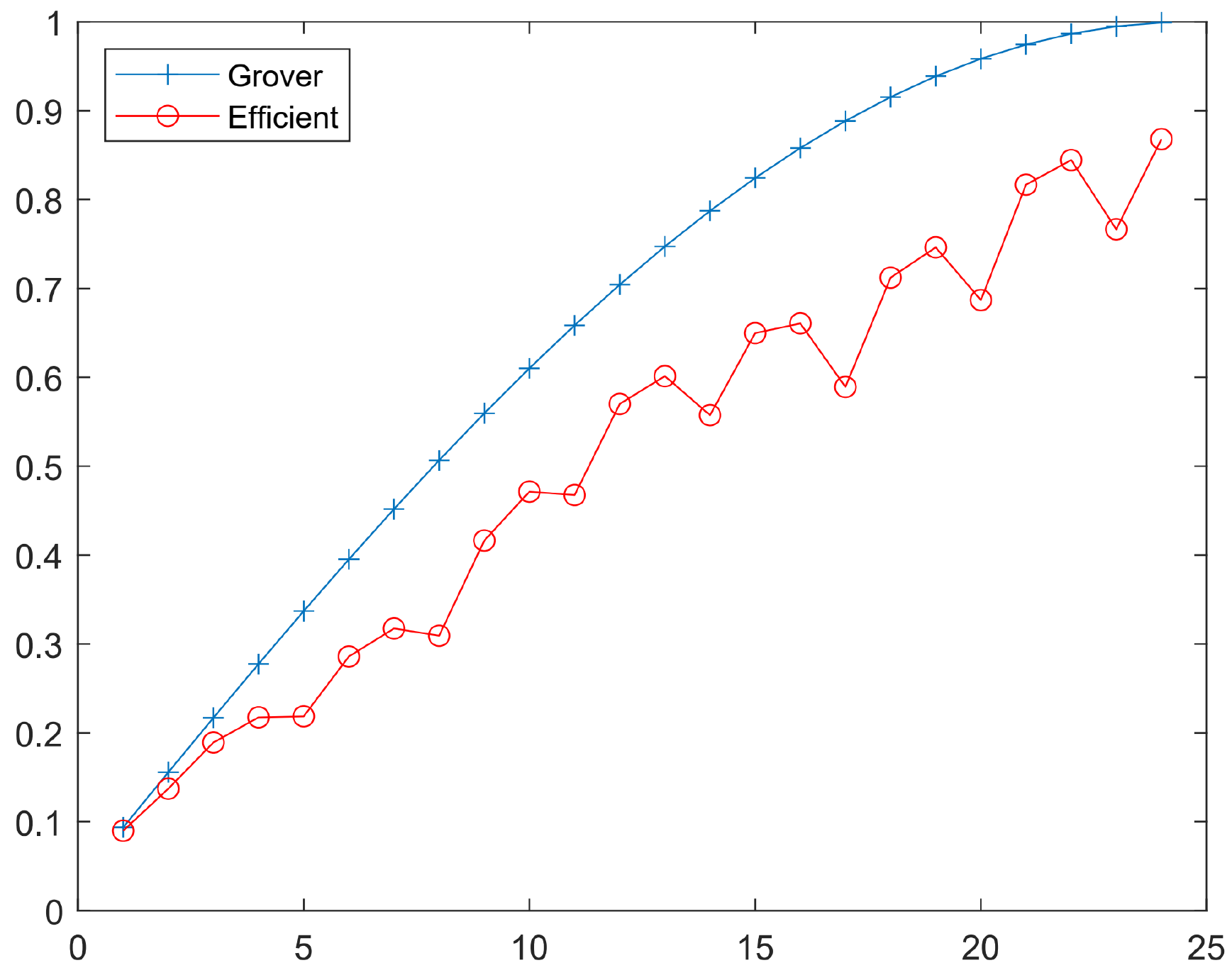}
    \caption{$n=10,m=5,k_1=k_2=3$}
    \label{subfig:grover_5_10_33}
    \end{subfigure}
    \begin{subfigure}[t]{0.4\textwidth}
    \centering
    \includegraphics[width=\columnwidth]{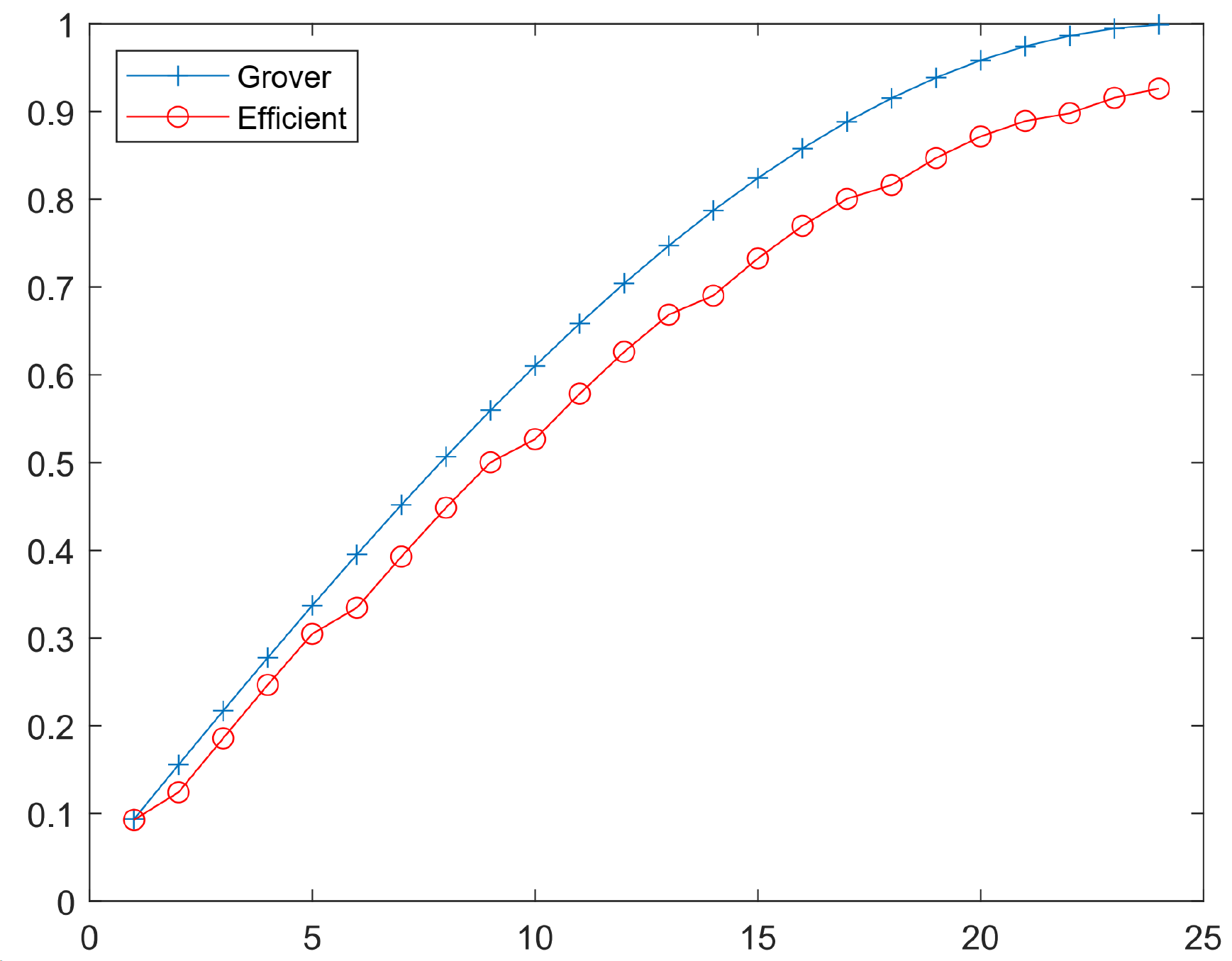}
    \caption{$n=10,m=3,k_1=k_2=1$}
    \label{subfig:grover_3_10_11}
    \end{subfigure}\hspace{2mm}
    \begin{subfigure}[t]{0.4\textwidth}{
    \centering
    \includegraphics[width=\columnwidth]{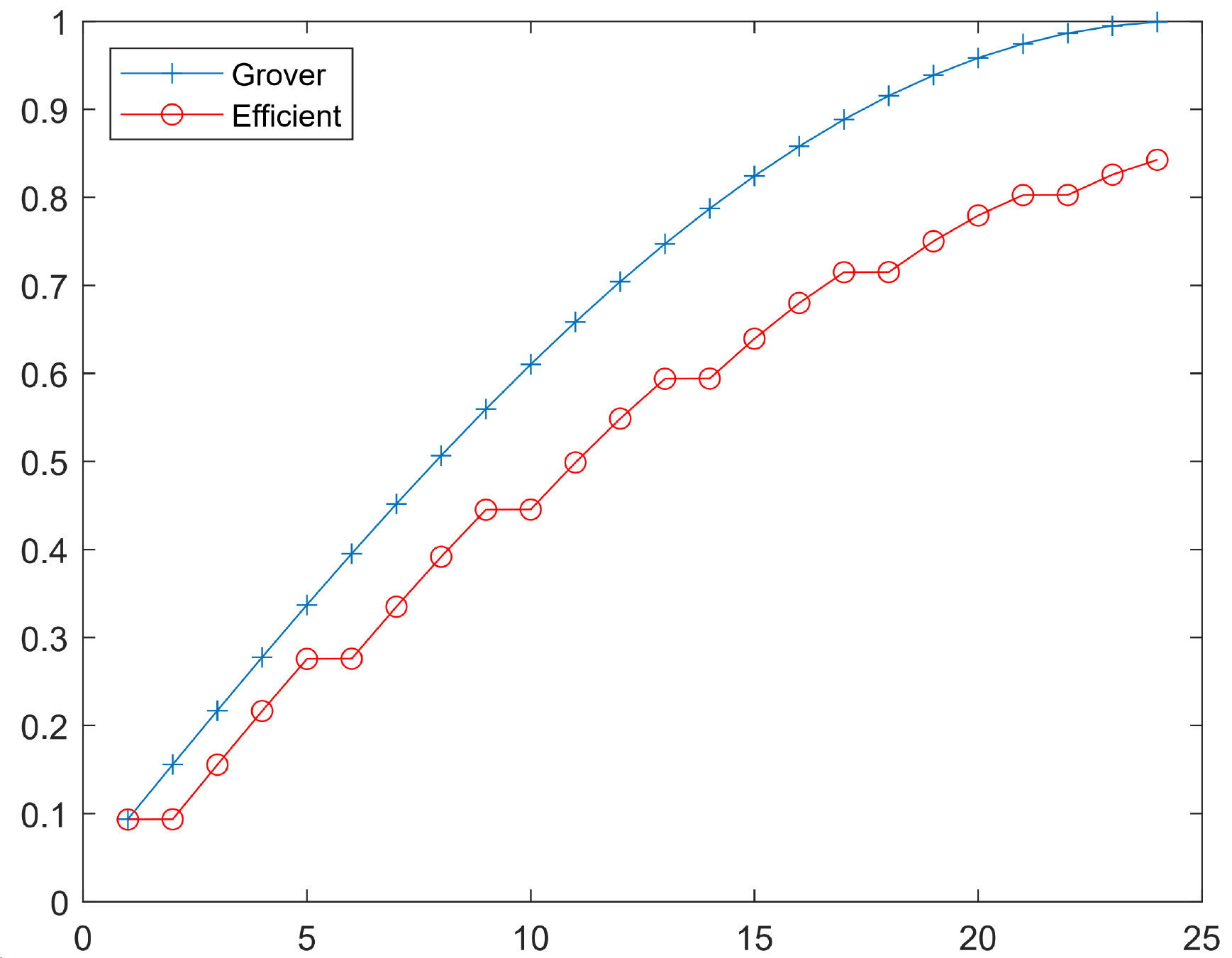}
    \caption{$n=10,m=2,k_1=k_2=1$}
    \label{subfig:grover_2_10_11}
    }
    \end{subfigure}
\caption{Four different designs of the 10-qubit hardware efficient quantum search algorithm}
\label{fig:efficient_different_designs}
\end{figure*}

The set of four normalized states $\ket{t},\ket{nt_m},\ket{nt_{n-m}},\ket{u}$ also form a basis. We name it the local search basis.
The transformation between the two eigenvectors basises and the local search basis can be represented with transition matrices $T_m$ and $T_{n-m}$.
\begin{equation}
\begin{pmatrix}
\ket{\psi_m^+}\\
\ket{\psi_m^-}\\
\ket{\psi_m^0}\\
\ket{\psi_m^1}
\end{pmatrix} = T_m\begin{pmatrix}
\ket{t}\\
\ket{nt_m}\\
\ket{nt_{n-m}}\\
\ket{u}
\end{pmatrix},\begin{pmatrix}
\ket{\psi_{n-m}^+}\\
\ket{\psi_{n-m}^-}\\
\ket{\psi_{n-m}^0}\\
\ket{\psi_{n-m}^1}
\end{pmatrix} = T_{n-m}\begin{pmatrix}
\ket{t}\\
\ket{nt_m}\\
\ket{nt_{n-m}}\\
\ket{u}
\end{pmatrix}
\end{equation}
\begin{equation}
T_{m} = \begin{pmatrix}
\frac{1}{\sqrt{2}}& -\frac{i}{\sqrt{2}}& 0&0\\
\frac{1}{\sqrt{2}} &\frac{i}{\sqrt{2}} &0 &0\\
0 & 0& sin(\theta) & cos(\theta)\\
0 & 0& cos(\theta)& -sin(\theta)
\end{pmatrix}
\end{equation}


\begin{equation}
T_{n-m} = \begin{pmatrix}
\frac{1}{\sqrt{2}}& 0& -\frac{i}{\sqrt{2}}&0\\
\frac{1}{\sqrt{2}} &0 &\frac{i}{\sqrt{2}} &0\\
0 & sin(\gamma) &0 & cos(\gamma)\\
0 & cos(\gamma)&0& -sin(\gamma)
\end{pmatrix}
\end{equation}

We can represent the local Grover operator to the power of $k_1$ in the local search basis through change of basis: $G^{k_1}_{m}=T^{-1}_{m}G'^{k_1}_{m}T_{m}$:
\begin{equation}
\resizebox{0.9\linewidth}{!}{$
\begin{pmatrix}
cos(2k\theta)& sin(2k\theta)&0 &0\\
-sin(2k\theta) & cos(2k\theta) & 0 &0\\
0 & 0 & sin^2(\theta) + (-1)^kcos^2(\theta) &sin(\theta)cos(\theta)(1-(-1)^k)\\
0 & 0 & sin(\theta)cos(\theta)(1-(-1)^k) & cos^2(\theta) + (-1)^ksin^2(\theta)
\end{pmatrix}$}
\end{equation}
Similarly, we have $G^{k_2}_{n-m} = T^{-1}_{n-m}G'^{k_2}_{n-m}T_{n-m}$:
\begin{equation}
\resizebox{0.9\linewidth}{!}{$
\begin{pmatrix}
cos(2k\gamma)& 0&sin(2k\gamma) &0\\
0 & sin^2(\gamma) + (-1)^kcos^2(\gamma) & 0 &sin(\gamma)cos(\gamma)(1-(-1)^k)\\
-sin(2k\gamma) & 0 & cos(2k\gamma) & 0\\
0 & sin(\gamma)cos(\gamma)(1-(-1)^k) & 0 & cos^2(\gamma) + (-1)^ksin^2(\gamma)
\end{pmatrix}$}
\end{equation}
Let $a(\theta,k)=sin^2(\theta) + (-1)^kcos^2(\theta),b(\theta,k) = sin(\theta)cos(\theta)(1-(-1)^k),c(\theta,k)=cos^2(\theta) + (-1)^ksin^2(\theta) $. We can compute the local search operator $G^{k_2}_{n-m}G^{k_1}_{m}$. The whole algorithm can be expressed as $G = (G^{k_2}_{n-m}G^{k_1}_{m})^k$:
\begin{equation}\label{eq:24}
\resizebox{0.9\linewidth}{!}{$
\begin{pmatrix}
cos(2k_2\gamma)cos(2k_1\theta)& cos(2k_2\gamma)sin(2k_1\theta)&sin(2k_2\gamma)a(\theta,k_1) &sin(2k_2\gamma)b(\theta,k_1)\\
-a(\gamma,k_2)sin(2k_1\theta) & a(\gamma,k_2)cos(2k_1\theta) & b(\gamma,k_2)b(\theta,k_1) &b(\gamma,k_2)c(\theta,k_1)\\
-sin(2k_2\gamma)cos(2k_1\theta) & -sin(2k_2\gamma)sin(2k_1\theta) & cos(2k_2\gamma)a(\theta,k_1) & cos(2k_2\gamma)b(\theta,k_1)\\
-b(\gamma,k_2)sin(2k_1\theta) & b(\gamma,k_2)cos(2k_1\theta) & c(\gamma,k_2)b(\theta,k_1) & c(\gamma,k_2)c(\theta,k_1)
\end{pmatrix}^k$}
\end{equation}

When $k_1 = k_2 = 1$, the matrix can be simplified to:
\begin{equation}
\resizebox{0.9\linewidth}{!}{$
\begin{pmatrix}
cos(2\gamma)cos(2\theta)& cos(2\gamma)sin(2\theta)& -sin(2\gamma)cos(2\theta) &sin(2\gamma)sin(2\theta)\\
cos(2\gamma)sin(2\theta) & -cos(2\gamma)cos(2\theta) & sin(2\gamma)sin(2\theta)) &sin(2\gamma)cos(2\theta)\\
-sin(2\gamma)cos(2\theta) & -sin(2\gamma)sin(2\theta) & -cos(2\gamma)cos(2\theta) & cos(2\gamma)sin(2\theta)\\
-sin(2\gamma)sin(2\theta) & sin(2\gamma)cos(2\theta) & cos(2\gamma)sin(2\theta) & cos(2\gamma)cos(2\theta)
\end{pmatrix}^k$}
\end{equation}

The final state is:
\begin{equation}
\label{eq:final}
\ket{\psi_{out}} = G\ket{\psi}
\end{equation}
where $G$ and $\ket{\psi}$ is defined in Equation~\ref{eq:24} and Equation~\ref{eq:7}.
Based on Equation~\ref{eq:final}, we can compute the probability of the target state $\ket{t}$.

\subsection{Design Discussions}
The number of local search $k_1$, and $k_2$ within each repetition will affect the efficiency of the algorithm. The total number of query is $k_{total} =(k_1 + k_2)k$. In order to reduce the total number of query, we need to justify the optimal sequence of $k_1$ and $k_2$. Actually, increasing the number of local search within each repetition will hinder the effectiveness of our algorithm. As shown in Figure~\ref{fig:effectoflocal}, applying the $G_m$ operator will flip the states in $S(u)$ and $S(nt_{n-m})$ about their average amplitude. Therefore, applying $G_m$ twice will not change the probability amplitude of the states in $S(u)$ and $S(nt_{n-m})$. When $k_1$ is odd number, the effect of amplifying the states in $S(nt_{n-m})$ is the same as applying $G_m$ operator once. To make the best use of amplitude amplification, $k_1$ and $k_2$ should both be 1. In Figure~\ref{fig:efficient_different_designs}, we show the probability amplitude growth of four different designs of the 10-qubit hardware efficient quantum search algorithm. As shown in Figure~\ref{subfig:grover_5_10_11} and~\ref{subfig:grover_5_10_33}, the increment of $k_1$ and $k_2$ leads to the decrement of the probability amplitude.  

Different partitions of the n qubits will also affect the efficiency of the algorithm. In the extreme case, when $m$ is small, the local search $G_m$ will not have enough states in $S(nt_m)$ to amplify the states in $S(t)$. The imbalance of $S(nt_m)$ and $S(nt_{n-m})$ might also hinder the efficiency of the algorithm. As shown in Figure~\ref{subfig:grover_5_10_11}, \ref{subfig:grover_3_10_11}, and~\ref{subfig:grover_2_10_11}, the closer to equal partition, the higher probability amplitude. In our following discussion we will assume $k_1 = k_2 = 1$, and $m \sim n/2$.

\subsection{Complexity}
In this section, we use the numerical calculation results to show that our algorithm has similar oracle complexity as Grover's algorithm. In Table~\ref{table:iterations}, we calculate the total number of iterations $k_{total}$ to reach $98\%$ probability with different algorithms. $diff$ is the difference between $k_{total}$ of the hardware efficient algorithm and the $k_{total}$ of Grover's algorithm. We compare three different partitions of the hardware efficient quantum search algorithm and Grover's algorithm. $E(m = \frac{n}{2})$ denotes the hardware efficient algorithm with $m = \frac{n}{2}$. In the experiments, $k_1$ and $k_2$ are set to 1. When $n$ is large, the result of $E(m = \frac{n}{2} - a)$ algorithm is the same as $E(m = \frac{n}{2} + a)$, we only show the results with $m<\frac{n}{2}$. N.A. means the algorithm with certain partition will never reach the $98\%$ probability.

\begin{table}[bhtp]
  \caption{The number of iterations to reach $98\%$ probability with different algorithms.}
  \label{table:iterations}
  \centering
  \small
    \resizebox{\columnwidth}{!}{%
  \begin{tabular}{|c|c|c|c|c|c|c|c|}
    \hline
     & Grover & \multicolumn{2}{c|}{$E(m=\frac{n}{2}-4)$} & \multicolumn{2}{c|}{$E(m=\frac{n}{2}-2)$} & \multicolumn{2}{c|}{$E(m=\frac{n}{2})$}\\
    \hline
    \hline
    $n$ qubits & $k_{total}$ & $k_{total}$ & $diff$ & $k_{total}$ & $diff$ & $k_{total}$ & $diff$\\
    \hline
    16 & 183 & N.A. & N.A. & 195 & 12 & 188 & 5\\
    \hline
    24 & 2927 & 2963 & 36 & 2936 & 9 & 2931 & 4\\
    \hline
    32 & 46834 & 46868 & 34 & 46843 & 9 & 46838 & 4\\
    \hline
    40 & 749342 & 749376 & 34 & 749351 & 9 & 749346 & 4\\
    \hline
    44 & 2997369 & 2997403 & 34 & 2997378 & 9 & 2997373 & 4\\
    \hline
    48 & 11989475 & 11989509 & 34 & 11989484 & 9 & 11989479 & 4\\
    \hline
  \end{tabular}
  }
\end{table}

From Table~\ref{table:iterations}, we can find that as the number of qubits increase, the total iteration difference between our hardware efficient algorithm and Grover's algorithm decreases. The results suggest that the oracle complexity only differs by a constant factor. Such findings indicate that when $m \sim \frac{n}{2}$, and $2^n \rightarrow \infty$, the oracle complexity of our hardware efficient algorithm approaches the complexity of Grover's algorithm: $k_{total} \sim\frac{\pi}{4}\sqrt{2^n}$.

\subsection{Depth Optimization}
Depth of the quantum circuit denotes the number of consecutive parallel gate operations. A quantum circuit with a smaller depth will have less execution time and less noise. The total depth of the circuit is the summation of the oracle depth and the diffusion operator depth: $d_{total} = \sum d_{O} + \sum d_{D}$.

The oracle is the circuit for a black-box function. Different databases will have different oracle realization. Our hardware efficient quantum search algorithm will not reduce $d_{O}$. Since our algorithm has the same oracle complexity as Grover's algorithm, $\sum d_{O}$ is the same for both algorithms. Our algorithm reduces the depth by optimizing the diffusion operator. We replace half of the global diffusion operators with local diffusion operator $D_m$, and the rest of them are replaced with local diffusion operator $D_{n-m}$. As shown in Figure~\ref{fig:naivegrovercircuit}, the depth of an n-qubit diffusion operator is the same as the depth of an n-qubit Toffoli gate. An n-qubit Toffoli gate can be linearly decomposed into the basic gates with one ancilla qubit~\cite{barenco1995toffolilinear}. Thus, $d_{D}\sim O(n)$. The depth optimization algorithm~\cite{zhang2020partialgroverdepth} based on partial search saved depth scales as $O(2^{n/2})$. On the other hand, the saved depth of our algorithm scales as $O(n2^{n/2})$. 

Another important metric is the minimum expected depth($MED$):
\begin{equation} 
MED = min\frac{d_{total}(j)}{P_n(j)}
\end{equation}

Here, $d_{total}(j)$ is the total depth with j iterations. $P_n(j)$ is the probability of the target state after j iterations.

\begin{figure*}[ht]
\centering 
    \begin{subfigure}[t]{0.32\textwidth}
        \includegraphics[width=\columnwidth]{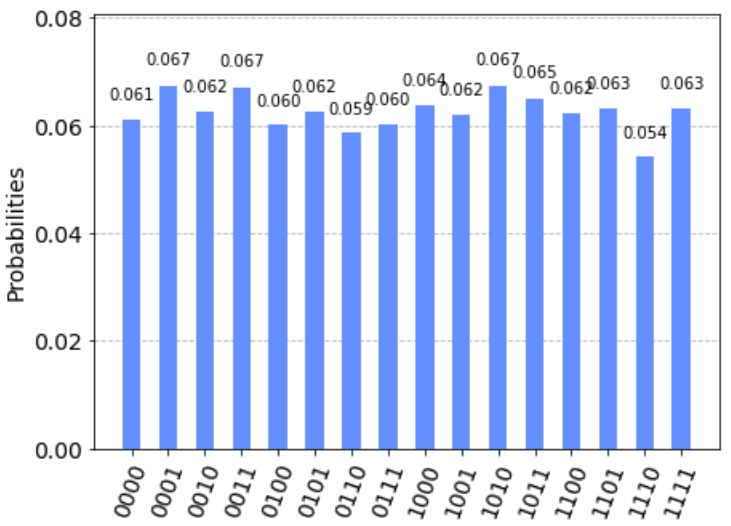}
        \caption{Grover's algorithm}
    \end{subfigure}\hfill
    \begin{subfigure}[t]{0.32\textwidth}
        \includegraphics[width=\columnwidth]{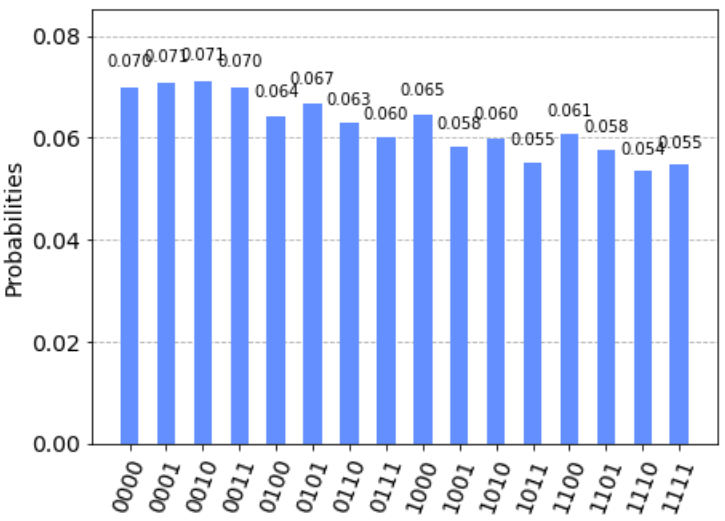}
        \caption{Partial Search Algorithm}
    \end{subfigure}\hfill
        \begin{subfigure}[t]{0.32\textwidth}
        \includegraphics[width=\columnwidth]{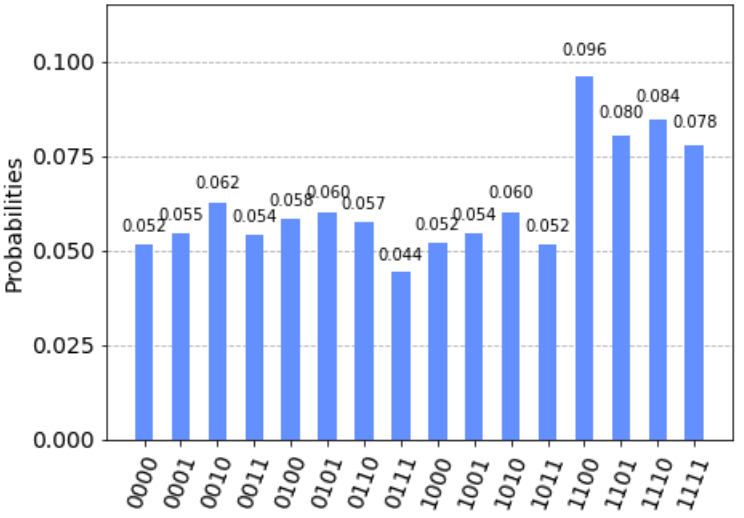}
        \caption{Hardware Efficient Search Algorithm}
    \end{subfigure}\hfill
    
\caption{Output distribution of three different 4-qubit search algorithms on `ibmq\_casablanca'}
\label{fig:output_distribution}
\end{figure*}

The prior work~\cite{zhang2020partialgroverdepth} shows their optimized partial search algorithm has lower $MED$ than the original Grover's algorithm. Here, we compare different algorithms in terms of the $MED$. The result is shown in Table~\ref{table:MED}. For Grover's search algorithm and the partial search algorithm, we use the optimal sequence described in prior work~\cite{zhang2020partialgroverdepth}. We also use the same diffusion operator decomposition method. The "G", "P", and "E" in the third row denotes "Grover algorithm", "Partial search algorithm", and our "Hardware efficient algorithm", respectively. As shown in Table~\ref{table:MED}, our algorithm has the smallest $MED$. Note the results in Table~\ref{table:MED} are the sequence of j Grover operators with the smallest $MED$. For example, in the six qubit case, four iterations result in the smallest $MED$. We can increase the number of iterations to increase the probability, but the $MED$ will also increase.
\begin{table}[bhtp]
  \caption{The MED of three 4-qubit quantum search algorithms.}
  \label{table:MED}
  \centering
  \small
    \resizebox{\columnwidth}{!}{%
  \begin{tabular}{|c|c|c|c|c|c|c|c|c|c|}
    \hline
     &  \multicolumn{3}{c|}{Probability} & \multicolumn{3}{c|}{$d_{total}$} & \multicolumn{3}{c|}{MED}\\
    \hline
    \hline
    n qubits & G & P & E & G & P & E & G & P & E\\
    \hline
    6 & 0.816 & 0.755 & 0.6318 & 504 & 360 & 272 & 617 & 477 & 431\\
    \hline
    7 & 0.833 & 0.887 & 0.757 & 1464 & 1173 & 949 & 1757 & 1323 & 1254\\
    \hline
    8 & 0.861 & 0.875 & 0.786 & 2916 & 2211 & 1758 & 3388 & 2527 & 2237\\
    \hline
    9 & 0.798 & 0.831 & 0.753 & 4848 & 3713 & 2959 & 6072 & 4470 & 3929\\
    \hline
    10 & 0.838 & 0.847 & 0.810 & 8712 & 6453 & 5225 & 10397 & 7614 & 6450\\
    \hline
  \end{tabular}
  }
\end{table}

\section{Methodology}
\label{sec:methodology}
Qiskit is an open-source quantum computing framework~\cite{Qiskit}. We run our experiments with Qiskit version 0.23. Our algorithm requires two local searches, which means the search algorithm needs at least four qubits. We run the 4-qubit Grover's search algorithm on a 7-qubit machine:`ibmq\_casablanca'. For the experiments with more than 4-qubits, we use the noise simulator from Qiskit Aer to simulate the noise effects. The noise simulator establishes a noise model based on the properties of actual hardware. In our experiment, we use the hardware properties of a 27-qubit machine:`ibmq\_toronto'. Each experiment is executed for 8192 shots. We use the highest Qiskit compiler optimization level associated with the relaxed peephole optimization~\cite{liu2020relaxed} to optimize the circuit and reduce the circuit depth. 


\section{Experimental Results}
\label{sec:results}
\subsection{Experiments on Real Quantum Computer}

We compare our hardware efficient algorithm with Grover's algorithm and partial search algorithm. We compare three 4-qubit quantum search algorithms on the 7-qubit machine: `ibmq\_casablanca'. Figure~\ref{fig:output_distribution} shows the output distribution of the search algorithms. The target state is $\ket{1100}$. The oracle consists of a multi-controlled Toffoli gate and several NOT gates. Grover's algorithm consists of three 4-qubit global search operators. The partial search algorithm consists of a local search for the first two qubits followed by a global search and another local search for the first two qubits. Our hardware efficient algorithm consists of a local search for the first two qubits, a local search for the last two qubits, followed by another local search for the first two qubits.

As shown in Figure~\ref{fig:output_distribution}, we can't infer the correct answer for the Grover's algorithm and the partial search algorithm. Due to the high noise-level, the output distribution is close to the equally distribution. The result is in accordance with the prior 4-qubit quantum search algorithm implementations\cite{mandviwalla2018grovernisq, wang2020grovernisq} on the ibmq machines where the success probability is approximately $6\%$. We can't infer the correct answer with the prior search algorithms. Compared to the Grover's algorithm and the partial search algorithm, our hardware efficient design leads to success rate improvement of $57\%$ and the correct result can be infered.  

The number of CNOT gates, circuit depth, and the success probability of these three search algorithms are shown in Table~\ref{table:gatecounts}. When mapping the logical qubits to the physical qubits, the compiler will introduce extra SWAP gates in the circuit. \textbf{\#CNOT} is the number of CNOT gates before qubit mapping, and \textbf{\#CNOT'} is the number of CNOT gates after mapping. \textbf{Depth} is the depth of the circuit before mapping, and \textbf{Depth'} is the depth of the circuit after mapping. \textbf{Prob} is the expected probability of the target item. \textbf{Prob'} is the actual probability of the target item. \textbf{IST} is the inference strength of the algorithm. It is defined as the ratio of the probability of the correct answer to the probability of the most frequently occurring wrong answer. The system can infer the correct answer when IST exceeds 1.

\begin{table}[bhtp]
  \caption{The \# CNOT gates, depth, success probability, and IST of three 4-qubit search algorithms on `ibmq\_casablanca'}
  \label{table:gatecounts}
  \centering
  \small
    \resizebox{\columnwidth}{!}{%
  \begin{tabular}{|c|c|c|c|c|c|c|c|}
    \hline
    & \textbf{\#CNOT} & \textbf{\#CNOT'} & \textbf{Depth} & \textbf{Depth'} & \textbf{Prob} & \textbf{Prob'} & \textbf{IST}\\
    \hline
    Grover & 78 & 141 & 154 & 192 & 0.96 & 0.062 & 0.93\\
    \hline
    Partial & 52 & 109 & 110 & 145 & 0.77 & 0.061 & 0.86 \\
    \hline
    Efficient & 33 & 79 & 76 & 112 & 0.77 & 0.096 & 1.14\\
    \hline
  \end{tabular}
  }
\end{table}

As shown in the table, our algorithm has the fewest CNOT gates and the least depth. The expected success probability of our algorithm is less than the others with the same number of queries. However, since our circuit has less noise, the final success probability is higher than the others. The experiment results show that even with less expected success probability, our algorithm still outperforms the other search algorithms on the NISQ computers.

\begin{table}[bhtp]
  \caption{The \#CNOT gates, depth, success probability, and IST of three search algorithms}
  \label{table:gatecountssimulator}
  \centering
  \small
    \resizebox{\columnwidth}{!}{%
  \begin{tabular}{|c|c|c|c|c|c|c|c|c|c|}
    \hline
     &  \multicolumn{3}{c|}{4-qubit} & \multicolumn{3}{c|}{5-qubit} & \multicolumn{3}{c|}{6-qubit}\\
    \hline
    metrics & G & P & E & G & P & E & G & P & E\\
    \hline
    \#CNOT & 72 & 50 & 33 & 103 & 82 & 50 & 139 & 115 & 76\\
    \hline
    \#CNOT' & 145 & 96 & 76 & 376 & 339 & 210 & 440 & 365 & 288\\
    \hline
    Depth & 151 & 113 & 79 & 204 & 168 & 115 & 276 & 228 & 156\\
    \hline
    Depth' & 358 & 222 & 166 & 584 & 472 & 329 & 681 & 577 & 473\\
    \hline
    Prob & 0.961 & 0.77 & 0.766 & 0.903 & 0.753 & 0.562 & 0.596 & 0.531 & 0.410\\
    \hline
    Prob' & 0.202 & 0.235 & 0.277 & 0.051 & 0.056 & 0.065 & 0.021 & 0.020 & 0.038\\
    \hline
    IST & 2.69 & 2.80 & 2.90 & 1.34 & 1.47 & 1.48 & 0.87 & 0.83 & 2.11\\
    \hline
  \end{tabular}
  }
\end{table}

\begin{figure*}[ht]
\centering 
    
    \begin{subfigure}[t]{0.32\textwidth}
        \includegraphics[width=\columnwidth]{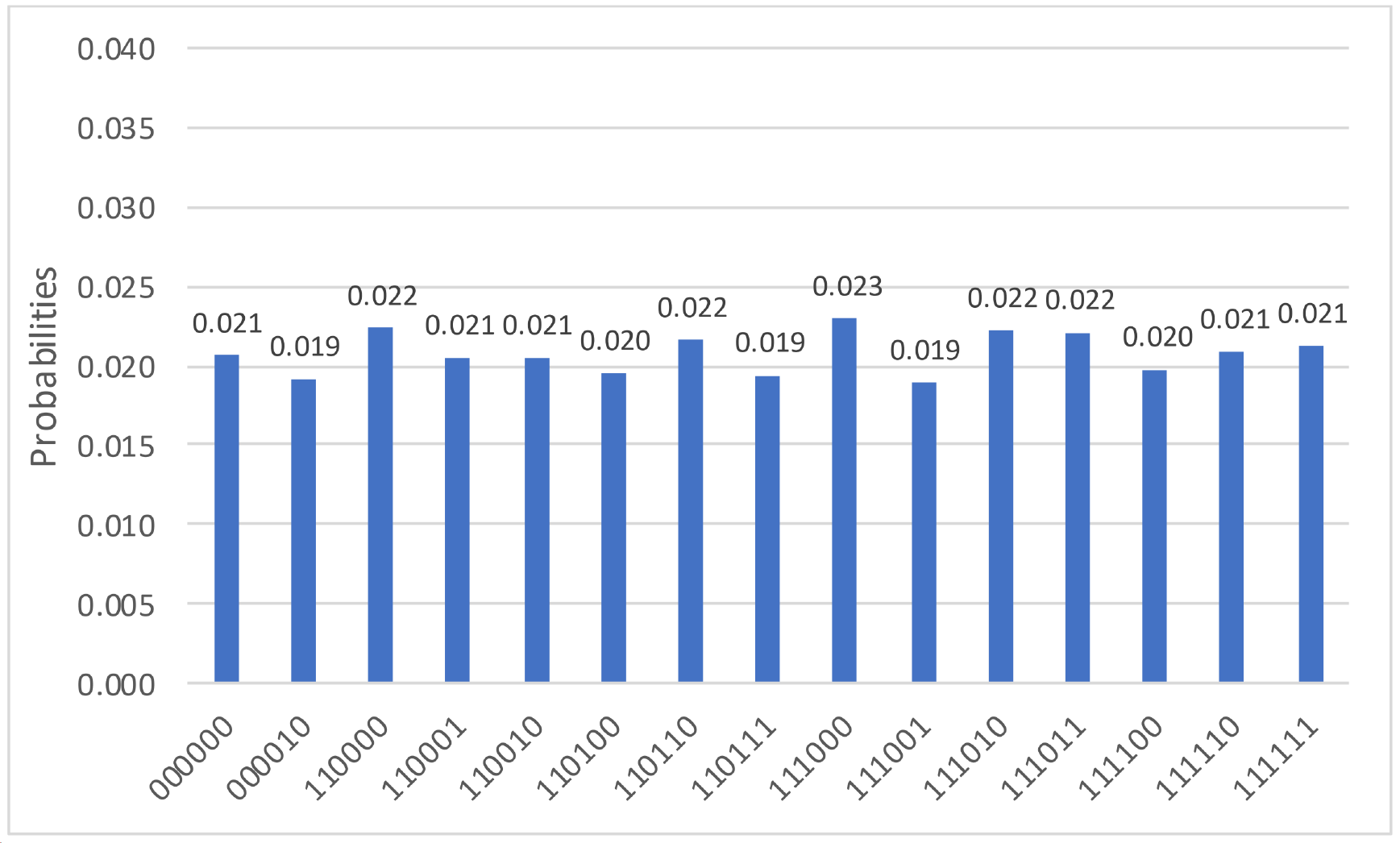}
        \caption{Grover's algorithm}
    \end{subfigure}\hfill
    \begin{subfigure}[t]{0.32\textwidth}
        \includegraphics[width=\columnwidth]{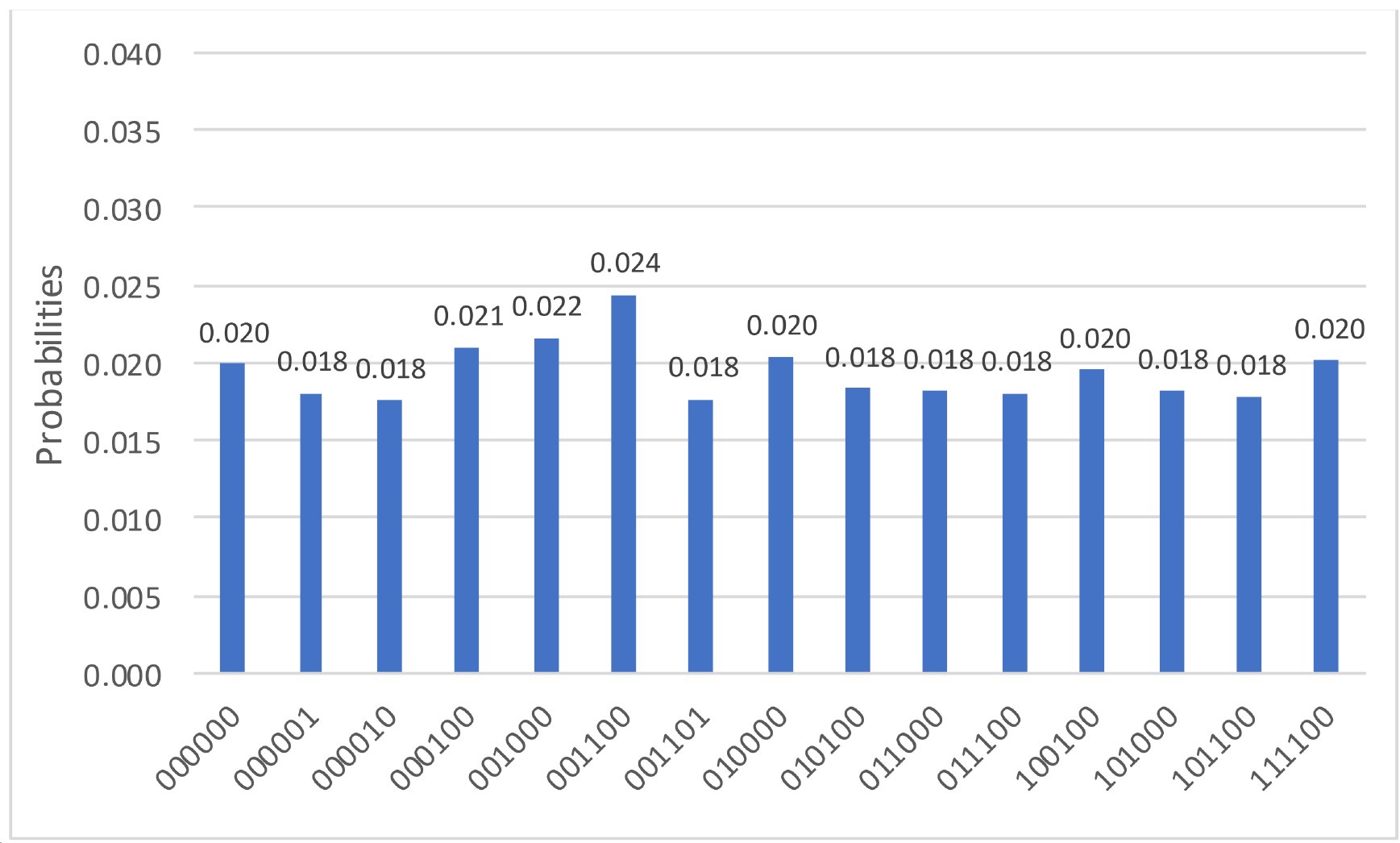}
        \caption{Partial Search Algorithm}
    \end{subfigure}\hfill
        \begin{subfigure}[t]{0.32\textwidth}
        \includegraphics[width=\columnwidth]{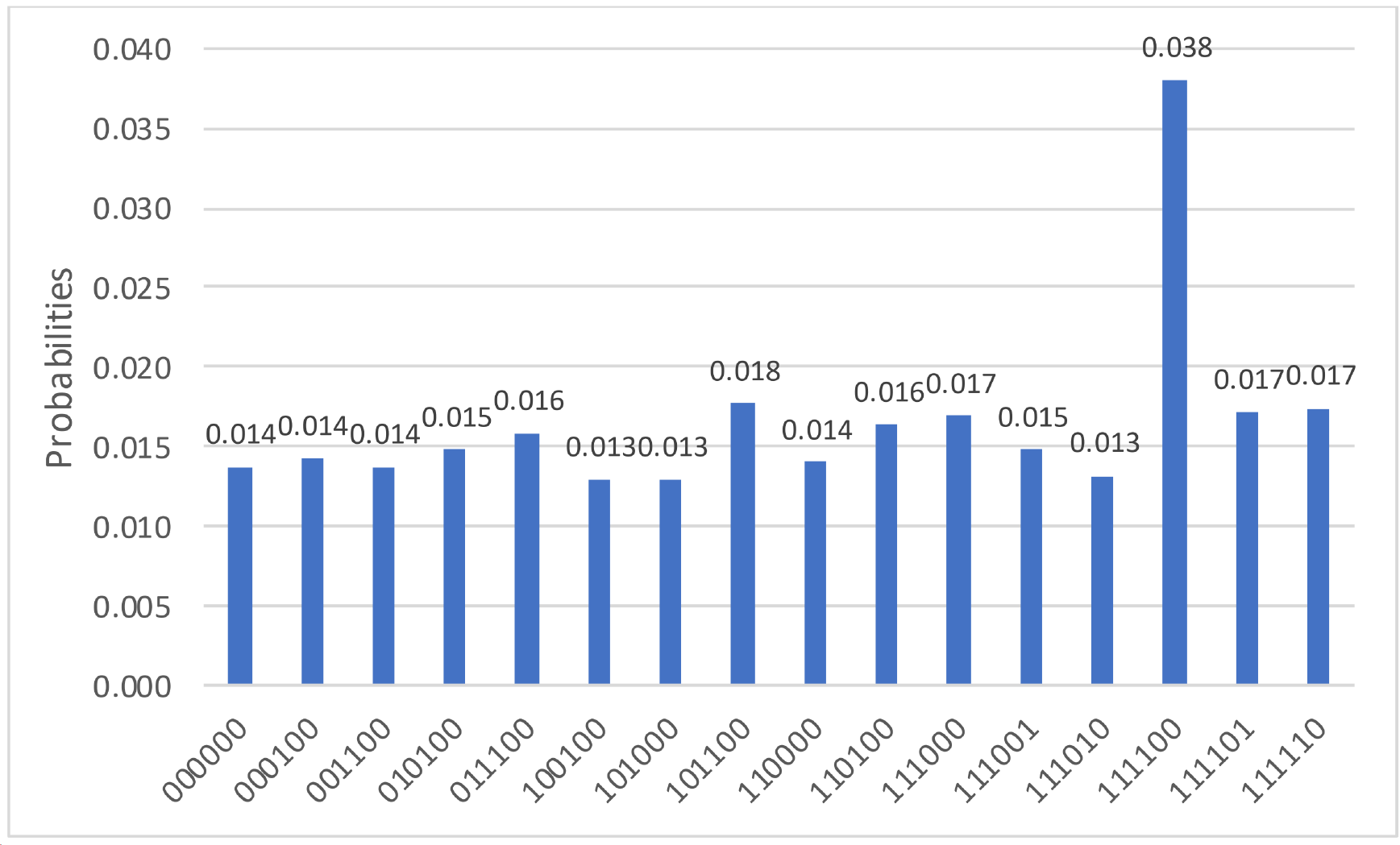}
        \caption{Hardware Efficient Search Algorithm}
    \end{subfigure}\hfill
\caption{Output distribution of three different 6-qubit search algorithms on `ibmq\_toronto' simulator}
\label{fig:output_distribution_simulator}
\end{figure*}

\subsection{Experiments on Noise Simulator}
We compare three quantum search algorithms on the noise simulator to show the scalability of our algorithm. We run the quantum search algorithm with qubit number from 4 to 6. Figure~\ref{fig:output_distribution} shows the output distribution of 6-qubit search algorithms. The target state is $\ket{111100}$. We only show the top 16 quantum states with high probability. Compared to the Grover's algorithm and the partial search algorithm, our hardware efficient design leads to success rate improvement of $90\%$. Grover's algorithm consists of three 6-qubit global search operators. The partial search algorithm consists of a local search for the first four qubits followed by a global search and another local search for the first four qubits. Our hardware efficient algorithm consists of a local search for the first three qubits, a local search for the last three qubits, followed by another local search for the first three qubits. 

The number of CNOT gates, circuit depth, and the success probability of these three search algorithms are shown in Table~\ref{table:gatecountssimulator}. Our algorithm achieves the fewest number of CNOT gates and least depth. The success rate of our algorithm is the highest among all three search algorithms.

\section{Conclusion}
In this paper, we propose a hardware efficient quantum search algorithm. The algorithm has similar oracle complexity as Grover's search algorithm, but the circuit depth is significantly reduced. We show that the quantum search can be achieved with only local diffusion operators. Our algorithm has less depth and minimum expected depth($MED$) than the prior work. Our experiments on the noise simulator highlight the effectiveness of our algorithm.
\label{sec:conclusion}

\bibliographystyle{ACM-Reference-Format}
\bibliography{ref}
                                                
\end{document}